\documentclass[superscriptaddress,secnumarabic,
amssymb,amsmath,nobibnotes,aps,prd,showkeys,showpacs,nofootinbib]{revtex4}%
\usepackage{graphicx}
\usepackage{epsf}
\usepackage{bm}
\usepackage{amsmath}
\usepackage{amsfonts}
\usepackage{amssymb}
\usepackage{epstopdf}
\usepackage{natbib}
\usepackage{color}
\usepackage{float}
\providecommand{\U}[1]{\protect\rule{.1in}{.1in}}

\newcommand{\be}{\begin{equation}}
\newcommand{\ee}{\end{equation}}

\newcommand{\mincir}{\raise
-3.truept\hbox{\rlap{\hbox{$\sim$}}\raise4.truept\hbox{$<$}\ }}
\newcommand{\magcir}{\raise
-3.truept\hbox{\rlap{\hbox{$\sim$}}\raise4.truept\hbox{$>$}\ }}

\newtheorem{remark}{Remark}[section]

\begin{document}

\title{An analytic formula to calculate the reheating temperature via gravitational particle production in smooth non-oscillating backgrounds }

\author{Jaume  de Haro}
\email{jaime.haro@upc.edu}
\affiliation{Departament de Matem\`atiques, Universitat Polit\`ecnica de Catalunya, Diagonal 647, 08028 Barcelona, Spain}

\author{Llibert Arest\'e Sal\'o}
\email{l.arestesalo@qmul.ac.uk}
\affiliation{School of Mathematical Sciences, Queen Mary University of London, Mile End Road, London, E1 4NS, United Kingdom}



\begin{abstract}
We present for smooth  non-oscillating backgrounds an analytic formula which calculates
 the energy density of  massive and massless  particles created via gravitational particle production, thus giving the corresponding reheating temperature. It can be applied to models of Quintessential Inflation such as $\alpha$-attractors, and shows that for masses larger than the Hubble rate at the end of inflation, namely $H_{END}$, the reheating temperature is exponentially suppressed. On the contrary, for masses of the order of $H_{END}$ one obtains a maximum reheating temperature of the order of $10^7$ GeV. Finally, to overcome the constraints coming from the overproduction of Gravitational Waves in Quintessential Inflation, we have shown that the viable masses which ensure the Big Bang Nucleosynthesis success are in the range between {$2\times 10^{10}$ GeV and $ 4\times 10^{13}$ GeV}, leading to a maximum reheating temperature of the order {$10^5-10^7$ GeV}. 
\end{abstract}

\vspace{0.5cm}

 \par
\maketitle


\section{Introduction}

The reheating mechanism is an essential ingredient  in inflationary cosmology \cite{guth} because the Universe needs to be reheated after its extreme growth in order to match with the hot Big Bang model. In Quintessential Inflation \cite{pv} (see also the recent papers \cite{hossain,geng,dimopoulos,dimopoulos1,Giovannini1}), which is one of the simplest theories which try to explain the early and late time accelerated expansion of the Universe, the most used reheating mechanisms are the {\it gravitational particle production} \cite{parker, Zeldovich, gmmbook} and the {\it Instant Preheating}. In both cases the numerical calculation of the energy density of the produced particles in viable models is very complicated or even impossible in practice. 
This is why  some toy models or approximations are considered
to obtain analytic formulas to calculate the energy density of the produced particles, and then, these analytic formulas are applied to more realistic models.  

\

More precisely, firstly, dealing with gravitational particle production of light particles non-conformally coupled to gravity, it was proved  that the energy density of the produced particles at the beginning of kination (the era after inflation where all the energy of the inflaton becomes kinetic) is of the order of $10^{-2} H_{kin}^4$ (see for instance \cite{ford, damour, giovannini}), being $H_{kin}$ the value of the Hubble rate at the beginning of kination, by using a toy model consisting of an abrupt and non-smooth phase transition from the de Sitter regime to the exact kination era. 
Then, for Instant Preheating the expansion of the universe is disregarded approximating the value of the scale factor with its value at the beginning of kination. Furthermore, the inflaton field is linearized around the beginning of kination, thus obtaining that the square of the  frequency of the vacuum modes is a quadratic function of the conformal time. For this kind of frequencies the $\beta$-Bogoliubov coefficient can be analytically computed and the number density of the produced particles at the beginning of kination is approximated by $\frac{(g\dot{\varphi}_{kin})^{3/2}}{8\pi^3}\exp\left(-\frac{\pi m_{\chi}^2}{g\dot{\varphi}_{kin}}\right)$ (see \cite{fkl0,fkl}), where $\varphi$ denotes the inflaton, the derivative is respect to the cosmic time, $m_{\chi}$ is the bare mass of the produced particles and $g$ is a dimensionless coupling constant between the inflaton and the quantum field that produces the particles.

\

For the same reason as explained above,   dealing with viable Quintessential Inflation models which come from a smooth potential  which leads to a non-oscillating background, 
the main goal of this work is to obtain a universal formula for the energy density of massive and massless particles gravitationally  produced during the phase transition from the end of the slow-roll phase to the beginning of kination. The idea for obtaining this formula is quite simple, we expand the scale factor up to order two around the end of inflation, which is the moment when the particles start getting produced, obtaining a quadratic approximation of the frequency of the vacuum modes.  Then, we use the well-known formula for quadratic frequencies to compute the $\beta$-Bogoliubov coefficient, which is the key ingredient to obtain the energy density of the produced particles and, thus, the reheating temperature of the Universe via gravitational particle production of massive and massless particles.

\

We have tested our analytic formula using an improved version of the toy model proposed in \cite{hashiba}, showing that our formula is in  agreement  with the numerical results.
In addition, we have also checked that the reheating via particle production of heavy particles overcomes the constraint that entails  the overproduction of Gravitational Waves in Quintessential Inflation.
Finally, we have applied it to calculate the reheating temperature for an exponential $\alpha$-attractor potential in the context of Quintessential Inflation.

\

Throughout  the manuscript we  use natural units, i.e., 
 $\hbar=c=k_B=1$,  and the reduced Planck's mass is denoted by $M_{pl}\equiv \frac{1}{\sqrt{8\pi G}}\cong 2.44\times 10^{18}$ GeV.

\section{Gravitational particle production in non-oscillating backgrounds }

 We will start this section studying the gravitational particle production of heavy massive particles conformally coupled to gravity. In this situation, 
the Klein-Gordon equation for the vacuum  modes is the one of a time-dependent harmonic oscillator,
\begin{eqnarray}\label{kg1}
\chi_{ k}''(\eta)+\omega^2_k(\eta) \chi_{ k}(\eta)=0,
\end{eqnarray}
where 
the prime denotes the derivative with respect to the conformal time $\eta$ and 
$\omega_k(\eta)$ is the time-dependent frequency, which
for conformally coupled particles
is given by 
\begin{eqnarray}\label{omega}
\omega_k^2(\eta)=k^2+a^2(\eta)m_{\chi}^2,\end{eqnarray}
being $m_{\chi}$ the mass of the produced particles and $a(\eta)$ the scale factor. 

\

{

In general, for non-oscillating smooth models it is impossible to find an exact analytic formula for the $\beta$-Bogoliubov coefficients. Therefore, we will do some approximation in order to find an analytic expression which agrees to good accuracy with the numerical calculation used to determine these coefficients. To do it,
first of all, on the one hand we have to take into account that real particles are produced during the phase transition from the end of inflation to the beginning of kination, and on the other hand that the $\beta$-Bogoliubov coefficients encodes the production of particles and also the vacuum polarization effects, which disappear soon after the beginning of kination when the Bogoliubov coeffients stabilize and then it also encodes the creation of particles. For this reason, the approximation will be better when we approximate the frequency during the phase transition and slightly after the beginning of kination, but it  does not matter what the approximation is like outside of this time period.

\

Therefore, taking into account this important fact, 
our idea is to approximate the scale factor around a time $\bar \eta$ by an expression of the form
\begin{eqnarray}
    a^2(\eta)\cong  A+ B(\eta-\bar\eta+C)^2,
\end{eqnarray}
where $A\geq 0$, $B>0$,  $C$ and $\bar\eta$ are some constants which we will determine right now, because for these backgrounds the Bogoliubov coefficients can be calculated analytically.

\

First of all we need to determine the time $\bar\eta$. Noting that the particle production occurs during the phase transition from the end of the slow-roll  to the beginning of kination, one can support that $\bar\eta$ has to be an instant when the universe is in this phase transition. So, we use the Taylor's expansion of the scale factor around $\bar\eta$ up to order two, obtaining
\begin{eqnarray}
    a(\eta)\cong \bar a+\bar a'(\eta-\bar\eta)+\frac{1}{2}\bar a''(\eta-\bar\eta)^2=\bar a+\bar a^2 \bar H(\eta-\bar\eta)+\frac{1}{12}\bar a^3 \bar R(\eta-\bar\eta)^2,
\end{eqnarray}
where we have introduced the notation $\bar a=a(\bar\eta)$,
$\bar H=H(\bar\eta)$ and $\bar R=R(\bar\eta)$, being $R$ the Ricci curvature. Thus, the square of the scale factor can be approximated by 
\begin{eqnarray}\label{no-sym}
    a^2(\eta)\cong \bar a^2+2\bar a^3 \bar H(\eta-\bar\eta)+\bar a^4 (3\bar{H}^2+\dot{\bar H})(\eta-\bar\eta)^2=\nonumber\\
    =\bar a^2+2\bar a^3 \bar H(\eta-\bar\eta)+
    \frac{3}{2}\bar a^4 \bar{H}^2(1-\bar w_{eff})(\eta-\bar\eta)^2 =\nonumber\\
    =\bar a^2\left(1-\frac{2}{3(1-\bar w_{eff})}\right)+
    \frac{3}{2}\bar a^4 \bar{H}^2(1-\bar w_{eff})\left(\eta-\bar\eta+\frac{2}{3\bar a\bar H (1-\bar w_{eff})}\right)^2     ,    
    \end{eqnarray}
where we have used the effective Equation of State parameter $w_{eff}=-1-\frac{2\dot{H}}{3H^2}$
and also the notation $\bar w_{eff}=w_{eff}(\bar\eta)$.

\

Now, since $A\geq 0$ and $B>0$, we can see that  
$\bar{w}_{eff}\leq  1/3$. Thus, inserting (\ref{no-sym}) in the frequency, one has 
\begin{eqnarray}
    \omega_k^2(\eta)\cong k^2+\bar a^2\left(1-\frac{2}{3(1-\bar w_{eff})}\right)
    m_{\chi}^2 +\frac{3}{2}\bar a^4 \bar{H}^2(1-\bar w_{eff})m_{\chi}^2
    \left(\eta-\bar\eta+\frac{2}{3\bar a\bar H (1-\bar w_{eff})}\right)^2,
\end{eqnarray}
and, defining 
\begin{eqnarray}
\tau\equiv\sqrt{\sqrt{\frac{3}{2}(1-\bar w_{eff})}\bar a^2\bar H m_{\chi}}
\left(\eta-\bar\eta+\frac{2}{3\bar a\bar H(1-\bar w_{eff})}\right),
\end{eqnarray}
the Klein-Gordon equation becomes
\begin{eqnarray}\label{kg2}
\frac{d^2\chi_k}{d\tau^2}+(\kappa^2+\tau^2)\chi_k=0,
\end{eqnarray}
where we have introduced the notation 
$\kappa^2=\frac{k^2+{\bar a^2\left(1-\frac{2}{3(1-\bar w_{eff})}  \right)
m_{\chi}^2}}{\sqrt{\frac{3}{2}(1-\bar w_{eff})}\bar a^2\bar H m_{\chi}
 }$.
Note that for this quadratic frequency the $\beta$-Bogoliubov coefficient is obtained using the well-known formula \cite{martin}
\begin{eqnarray}\label{beta2}
       |\beta_k|^2= e^{-\pi\kappa^2}=
       \exp\left(-\pi\frac{k^2+{\bar a^2\left(1-\frac{2}{3(1-\bar w_{eff})}  \right)
       m_{\chi}^2}}{\sqrt{\frac{3}{2}(1-\bar w_{eff})        }\bar a^2\bar H m_{\chi}  }\right),
   \end{eqnarray}}
which can be derived as follows.
First, recall  that the positive frequency modes in the WKB approximation are
\begin{eqnarray}
\phi_{k,+}(\tau)=\frac{1}{(\kappa^2+\tau^2)^{1/4}}e^{-i\int \sqrt{\kappa^2+\tau^2}d\tau},
\end{eqnarray}
and for large values of $|\tau|$ ($|\tau|\gg \kappa$)  one can make the approximations
$(\kappa^2+\tau^2)^{1/4}\cong |\tau|^{1/2}$ and $\sqrt{\kappa^2+\tau^2}\cong |\tau|\left(1+\frac{\kappa^2}{2\tau^2}  \right)$, obtaining
\begin{eqnarray}
\phi_{k,+}(\tau\ll -\kappa)\cong |\tau|^{-1/2+i\kappa^2/2}e^{i \tau^2/2}, \qquad \phi_{k,+}(\tau\gg \kappa)\cong |\tau|^{-1/2-i\kappa^2/2}e^{-i\tau^2/2},
\end{eqnarray}
while for the negative frequency modes 
\begin{eqnarray}
\phi_{k,-}(\tau\gg \kappa)=\frac{1}{(\kappa^2+\tau^2)^{1/4}}e^{i\int \sqrt{\kappa^2+\tau^2}d\tau}\cong |\tau|^{-1/2+i\kappa^2/2}e^{i \tau^2/2}.
\end{eqnarray}
On the other hand,  the positive frequency  modes evolve as 
\begin{eqnarray}\label{evolution}
\phi_{k,+}(\tau\ll -\kappa)\longrightarrow \alpha_k\phi_{k,+}(\tau\gg \kappa)+\beta_k\phi_{k,-}(\tau\gg \kappa).
\end{eqnarray}
So, to calculate the Bogoliubov coefficients one can use the WKB method in the complex plane integrating the frequency along the path $\gamma=\{ z=|\tau|e^{i\alpha}, -\pi \leq \alpha\leq 0\}$, obtaining that for $\tau\gg \kappa$ the early time positive frequency modes evolve at late time as
\begin{eqnarray}
e^{-\frac{\kappa^2}{2}\pi}|\tau|^{-1/2+i\kappa^2/2}e^{i\tau^2/2},
\end{eqnarray}
and comparing with (\ref{evolution}) one gets
\begin{eqnarray}
|\beta_k|^2\cong e^{-\kappa^2\pi} \qquad \mbox{and}\qquad |\alpha_k|^2=1+|\beta_k|^2=1+e^{-\kappa^2\pi},
\end{eqnarray}
and,  since $\omega_k(\eta)\cong a(\eta)m_{\chi}$,  the energy density of the massive produced particles can be approximated during the kination by
{
\begin{eqnarray}\label{energy}
    \langle \rho(\eta)\rangle\cong \frac{1}{8\pi^3}
    \left( \frac{3}{2}(1-\bar w_{eff}) \right)^{3/4}
     \exp\left(-\pi 
    \frac{\sqrt{2}(1-3\bar w_{eff})m_{\chi}
    }{3\sqrt{{3}(1-\bar w_{eff})^3}\bar H} \right)     \sqrt{\frac{m_{\chi}}{\bar H}}m_{\chi}^2\bar H^2
   \left( \frac{\bar a}{a(\eta)} \right)^3,
\end{eqnarray}
where here it is important to point out that this formula is universal in the sense that it is independent of the smooth non-oscillating model.

\

At this point,  one has to choose a reasonable value of $\bar\eta$.  Taking:
\begin{enumerate}
   \item $\bar\eta=\eta_{END}$, where $``END"$ denotes the end of inflation. Taking into account that at the end of the early accelerated expansion one has $\bar w_{eff}= -1/3$, the energy density of the produced particles is given by 
\begin{eqnarray}\label{rho0}
    \langle \rho(\eta)\rangle\cong \frac{1}{4\pi^3} e^ {- \frac{\pi m_{\chi}}{2\sqrt{2} H_{END}} }    
    \sqrt{\frac{m_{\chi}}{ \sqrt{2} H_{END}}}m_{\chi}^2 H^2_{END}
     \left( \frac{ a_{END}}{a(\eta)} \right)^3.
\end{eqnarray}

\item $\bar\eta=\eta_m$, where $\eta_m$ means the instant, during the phase transition,  when $\bar w_{eff}=0$. In this case one has 
\begin{eqnarray}
    \langle \rho(\eta)\rangle\cong \frac{3}{16\pi^3} e^{- \frac{\sqrt{2}\pi m_{\chi}}{3\sqrt{3} H_m} }     \sqrt{ \sqrt{\frac{2}{3}}\frac{m_{\chi}}{ H_m}}m_{\chi}^2 H^2_m
    \left( \frac{ a_m}{a(\eta)} \right)^3.
\end{eqnarray}

\item $\bar\eta=\eta_r$, where $\eta_r$ means the instant, during the phase transition,  when $\bar w_{eff}=1/3$, which leads to
\begin{eqnarray}
    \langle \rho(\eta)\rangle\cong \frac{1}{8\pi^3}
    \sqrt{\frac{m_{\chi}}{ H_r}}m_{\chi}^2 H^2_r
     \left( \frac{ a_r}{a(\eta)} \right)^3.
\end{eqnarray}

\end{enumerate}

\

\begin{figure}[H]
    \centering
    \includegraphics[width=300pt]{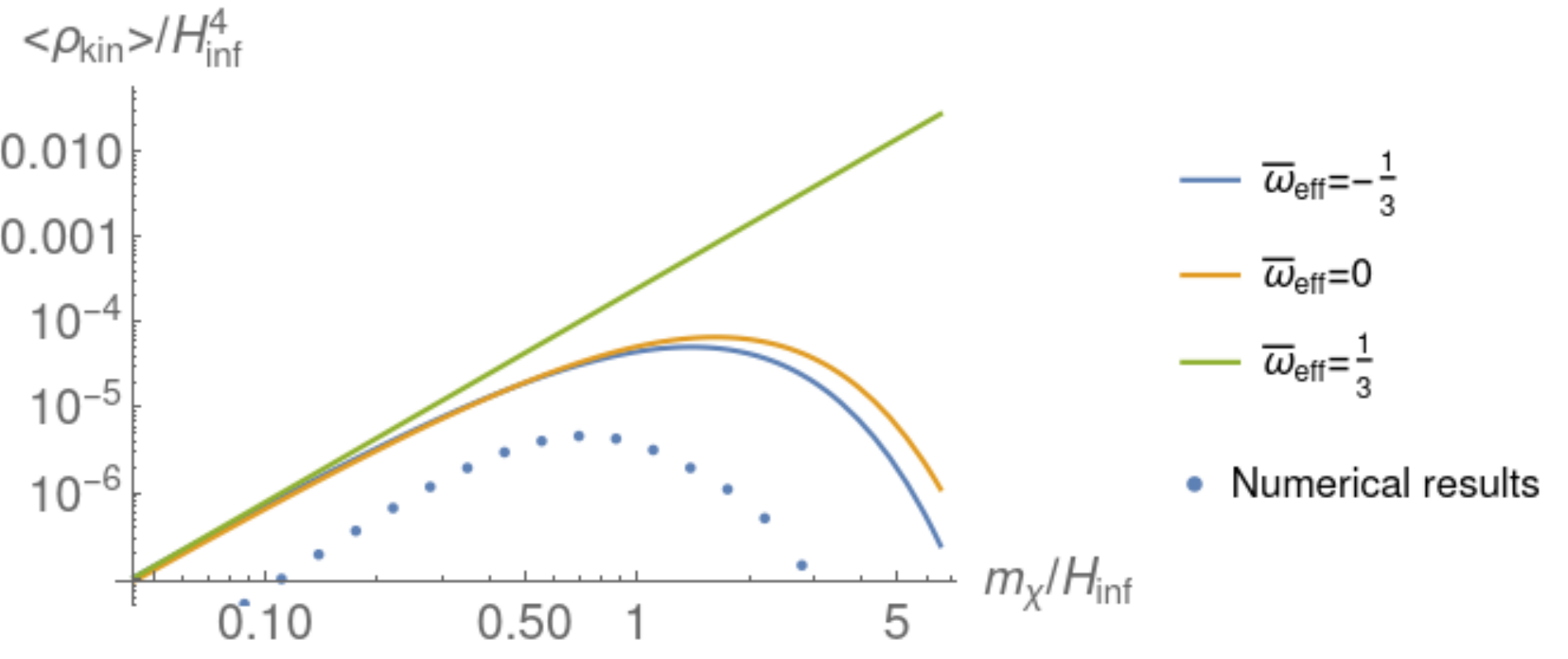}
    \caption{Numerical and analytical calculation of the energy density of produced particles  at the beginning of kination, as a function of the mass for the model (\ref{background}). Using 
      $H_{inf}= 10^{-6} M_{pl}$ and $\Delta\eta=\frac{1}{2H_{inf}}$. }
    \label{fig:my_label10}
\end{figure}

\

In Figure \ref{fig:my_label10} we have tested the analytic formulae for the toy model (\ref{background}) depicted in the Section \ref{toy-model} and  we can see that the energy density which matches the best with the numerical results is when one chooses $\bar\eta=\eta_{END}$.

\

In addition, with $\bar\eta=\eta_{END}$, by using the approximation $a_{END}\cong a_{kin}$,  the energy density at the kination becomes
\begin{eqnarray}\label{rho1}
   \langle \rho_{kin}\rangle
   \cong
   \frac{1}{4\pi^3}
   e^{-\frac{\pi m_{\chi}}{2\sqrt{2}H_{END}}}
   \sqrt{\frac{m_{\chi}}{\sqrt{2}H_{END}}} H_{END}^2m_{\chi}^2, \end{eqnarray}
showing  that the energy density of the produced particles is exponentially suppressed for masses larger than $H_{END}$. In addition, it  
only depends on the mass  and the value 
of the Hubble rate at the end of inflation, which can be computed analytically for a given potential using that
$H_{END}=\sqrt{\frac{V(\varphi_{END})}{2M_{pl}^2}}$, where $\varphi_{END}$ is calculated at the moment when the slow-roll parameter $\epsilon\equiv\frac{M_{pl}^2}{2}\left(\frac{V_{\varphi}}{V} \right)^2$ is equal to $1$.

\

Finally, 
  taking into account that in the majority of models inflation ends when  $H_{END}\sim 10^{-6} M_{pl}$,
one arrives at the approximation
\begin{eqnarray}
   \langle \rho_{kin}\rangle
   \cong
   8\times 10^{-12}
   \exp\left({-\frac{5\pi\times 10^5 m_{\chi}}{\sqrt{2}M_{pl}}}\right)
   \sqrt{\frac{m_{\chi}}{\sqrt{2}M_{pl}}} M_{pl}^2m_{\chi}^2, \end{eqnarray}
which only depends on the mass of the produced particles.

\

}

\subsection{Non-conformally coupled particles}

When the quantum field is non-conformally coupled to gravity, the vacuum modes satisfy the Klein-Gordon equation
\begin{eqnarray}\label{kg2}
\chi_{ k}''(\eta)+\Omega^2_k(\eta) \chi_{ k}(\eta)=0,
\end{eqnarray}
where 
\begin{eqnarray}\Omega_k^2(\eta)=k^2+a^2(\eta)m_{\chi}^2+\left(\xi-\frac{1}{6}\right)a^2(\eta) R(\eta),\end{eqnarray}
being $\xi$ the coupling constant,  which we choose for convenience  to be $\xi\leq1/6$,  and $R(\eta)$  is the Ricci scalar. 

\

In addition, for the non-conformally coupled case, the vacuum energy density is given by \cite{Bunch}
\begin{eqnarray}\label{bunch}
    \langle \rho\rangle= \frac{1}{4\pi^2 a^4}
    \int_0^{\infty}dk~k^2\left\{|\chi_k'|^2+(\omega_k^2+
    (1-6\xi)a^2H^2)|\chi_k|^2+(6\xi-1)aH (\chi_k\bar{\chi}_k'+\chi_k'\bar{\chi}_k)   \right\}.
\end{eqnarray}

Therefore, using the diagonalization method 
 \cite{Zeldovich} (see also Section $9.2$ of \cite{gmmbook}),  we will write
the modes as follows,
\begin{eqnarray}\label{zs}
\chi_{k}(\eta)= \alpha_k(\eta)\phi_k(\eta)+
\beta_k(\eta)\bar{\phi}_k(\eta),\end{eqnarray}
where 
we have introduced the notation
\begin{eqnarray}\label{modes}
\phi_k(\eta)= \frac{e^{-i\int^{\eta} \Omega_k(\tau)d\tau}}{\sqrt{2\Omega_k(\eta)}},\qquad 
\bar{\phi}_k(\eta)= \frac{e^{i\int^{\eta} \Omega_k(\tau)d\tau}}{\sqrt{2\Omega_k(\eta)}},
\end{eqnarray}
and  we have to impose that the modes satisfy   the condition
\begin{eqnarray}
\chi_{k}'(\eta)= -i\Omega_k(\eta)\left(\alpha_k(\eta)\phi_k(\eta)-
\beta_k(\eta)\bar{\phi}_k(\eta)\right),\end{eqnarray}
obtaining that 
 the time-dependent Bogoliubov coefficients must satisfy the system 
\begin{eqnarray}\label{Bogoliubovequation1}
\left\{ \begin{array}{ccc}
\alpha_k'(\eta) &=& {\Omega_k'(\eta)}
\bar{\phi}_k^2(\eta)\beta_k(\eta)\\
\beta_k'(\eta) &=& {\Omega_k'(\eta)}\phi_k^2(\eta)\alpha_k(\eta).\end{array}\right.
\end{eqnarray}

\

\begin{remark}
    Dealing with   massless particles nearly conformally coupled to gravity, 
    one can treat the term $\left(\xi-\frac{1}{6}\right) a^2(\eta)R(\eta)$ as a perturbation. Then, we have  $\Omega_k(\eta)\cong k + \frac{1}{2k}\left(\xi-\frac{1}{6}\right) a^2(\eta)R(\eta)$ and, using the justified approximation $\alpha_k(\eta)\cong 1$, the equation \eqref{Bogoliubovequation1} becomes 
\begin{eqnarray}
    \beta_k'(\eta)\cong \frac{1}{4k^2}\left(\xi-\frac{1}{6}\right) (a^2(\eta)R(\eta))' e^{-2ik\eta},\end{eqnarray}
    whose solution, after integration by parts,  is 
    \begin{eqnarray}
    \beta_k  \cong\frac{1}{4k^2}\left(\xi-\frac{1}{6}\right) \int_{-\infty}^{\infty}(a^2(\eta)R(\eta))' e^{-2ik\eta}d\eta=
    \frac{i}{2k}\left(\xi-\frac{1}{6}\right) \int_{-\infty}^{\infty}a^2(\eta)R(\eta)e^{-2ik\eta}d\eta,
    \end{eqnarray}
which reproduces the well-known  perturbative result obtained for the first time in \cite{starobinsky} (see also \cite{ford}).
\end{remark}

\

Coming back to the vacuum energy density (\ref{bunch}),  a simple calculation shows that
\begin{eqnarray}
\chi_k\bar{\chi}_k'+\chi_k'\bar{\chi}_k=2i\Omega_k(
\bar{\alpha}_k\beta_k\bar{\phi}_k^2-{\alpha}_k\bar{\beta}_k{\phi}_k^2)=2i\frac{\Omega_k}{\Omega_k'}
(\bar{\alpha}_k\alpha_k'-\bar{\beta}_k\beta_k')
=2i\frac{\Omega_k}{\Omega_k'}
({\beta}_k\bar{\beta}_k'-\bar{\beta}_k\beta_k'),\end{eqnarray}
meaning that  this term vanishes when the Bogoliubov coefficient stabilizes, which happens soon after the beginning of kination. So, one can safely conclude that this term only contains vacuum polarization effects and, thus, it  does not contribute to the particle production.

\

Bearing this in mind,    at the beginning of kination the energy density of the produced particles is 
\begin{eqnarray}
    \langle \rho_{kin}\rangle= \frac{1}{4\pi^2 a^4_{kin}}
    \int_0^{\infty}dk k^2\left\{|\chi_k'|^2+\Omega_{k,kin}^2|\chi_k|^2   \right\}=
    \frac{1}{2\pi^2 a^4_{kin}}
    \int_0^{\infty}dk k^2\Omega_{k,kin}|\beta_k|^2 \cong
    \frac{m_{\chi}}{2\pi^2a_{kin}^3}    
    \int_0^{\infty}k^2|\beta_k|^2dk, 
\end{eqnarray}
where we have used the notation $\Omega_{k,kin}=\Omega_{k}(\eta_{kin})$,  that  at the kination 
$R_{kin}=-6H_{kin}^2$, and we have considered heavy massive particles 
because in this case one can make the approximation $\Omega_{k,kin}\cong 
m_{\chi}a_{kin}$.

\

On the other hand, to calculate the $\beta$-Bogoliubov coefficient, we make the same 
quadratic approximation for the symmetric scale factor at the end of the inflation as above, but for the Ricci scalar we consider its value 
at the beginning of kination, i.e. when $\dot{H}=-3H^2$, that is $R_{kin}\cong -6H_{kin}^2$. Thus,  we have
{
\begin{eqnarray}
\Omega_k^2(\eta)
\cong  k^2 +\frac{a^2_{END}}{2}(m_{\chi}^2+(1-6\xi)H_{kin}^2)
+2
a_{END}^4H_{END}^2(m_{\chi}^2+(1-6\xi)H_{kin}^2)\left(\eta-\eta_{END}+\frac{1}{2a_{END}H_{END}} \right)^2.\end{eqnarray}
Consequently the $\beta$-Bogoliubov is given by
\begin{eqnarray}
|\beta_k|^2=\exp\left(-\frac{\pi(k^2+\frac{a^2_{END}}{2}
(m^2_{\chi}+(1-6\xi)H_{kin}^2) )}
{\sqrt{2}a^2_{END}\sqrt{m^2_{\chi}+(1-6\xi)H_{kin}^2}H_{END}}\right),
   \end{eqnarray}
and the energy density of the produced particles at the beginning of kination is 
\begin{eqnarray}\label{master}
    \langle \rho_{kin}\rangle \cong   
   \frac{1}{4\pi^3}
   e^{-\frac{\pi \sqrt{m_{\chi}^2+(1-6\xi)H_{kin}^2}}{2\sqrt{2}H_{END}}}
   \sqrt{\frac{\sqrt{m_{\chi}^2+(1-6\xi)H_{kin}^2}}{\sqrt{2}H_{END}}} H_{END}^2 m_{\chi}\sqrt{m_{\chi}^2+(1-6\xi)H_{kin}^2}
   \left(\frac{a_{END}}{a_{kin}} \right)^3
\end{eqnarray}
and, since during the phase transition the scale factor remains nearly constant and there is no substantial drop of energy $H_{kin}\cong H_{END}$, one gets
\begin{eqnarray}\label{master1}
    \langle \rho_{kin}\rangle \cong   
   \frac{1}{4\pi^3}
   e^{-\frac{\pi \sqrt{m_{\chi}^2+(1-6\xi)H_{END}^2}}{2\sqrt{2}H_{END}}}
   \sqrt{\frac{\sqrt{m_{\chi}^2+(1-6\xi)H_{END}^2}}{\sqrt{2}H_{END}}} H_{END}^2 m_{\chi}\sqrt{m_{\chi}^2+(1-6\xi)H_{END}^2},
\end{eqnarray}
which only depends on the mass, the coupling $\xi$ and the value 
of the Hubble rate at the end of inflation.

\

On the other hand, for light particles, one can make the approximation $\Omega_k(\eta_{kin})\cong k$
and, disregarding the drop of energy during the phase transition,  the energy density is given by 
\begin{eqnarray}\label{master2}
    \langle \rho_{kin}\rangle \cong   
   \frac{1}{8\pi^3\sqrt{\pi}}
   e^{-\frac{\pi \sqrt{m_{\chi}^2+(1-6\xi)H_{END}^2}}{2\sqrt{2}H_{END}}} 
   H_{END}^2 ({m_{\chi}^2+(1-6\xi)H_{END}^2}).
\end{eqnarray}

\

In addition, 
for very light particles minimally coupled to gravity,   we have 
\begin{eqnarray}\label{GW}    
\langle \rho_{kin}\rangle \cong   
   \frac{1}{8\pi^3\sqrt{\pi}}
    H_{END}^4\cong 2\times 10^{-3} H_{END}^4, 
\end{eqnarray}
which agrees very well with the previous results \cite{damour, giovannini}.

\

Finally,  in the massless minimally coupled case we can easily calculate the reheating temperature. Effectively, denoting by ``end'' the end of kination, which occurs when the energy density of the produced particles is of the same order as the energy density of the inflaton field, and recalling that during the kination phase the energy density of the produced particles scales as $a^{-4}$ and the one of the inflaton as $a^{-6}$, one has that at the end of the kination
\begin{eqnarray}
    \langle \rho_{kin}\rangle \left(\frac{a_{kin}}{a_{end}} \right)^4\cong  3M_{pl}^2H_{kin}^2\left(\frac{a_{kin}}{a_{end}} \right)^6,\end{eqnarray}
thus implying that $\left(\frac{a_{kin}}{a_{end}} \right)^2\cong \frac{ \langle \rho_{kin}\rangle} {3M_{pl}^2H_{kin}^2}$. Hence, at the end of kination the energy density of the produced particles is given by
$ \langle \rho_{end }\rangle\cong \frac{\langle \rho_{kin}\rangle^3}{9M_{pl}^4H_{kin}^4} \cong \frac{\langle \rho_{kin}\rangle^3}{9M_{pl}^4H_{END}^4}$, and from the Boltzmann-Stefan law the reheating temperature is given by
\begin{eqnarray}\label{reh}
    T_{\text{reh}}=\left(\frac{30}{g_{reh}\pi^2} \right)^{1/4}\langle \rho_{end }\rangle^{1/4}
    \cong \left(\frac{10}{3g_{reh}\pi^2} \right)^{1/4}  
    \frac{\langle \rho_{kin }\rangle^{3/4}}{M_{pl} H_{END}}  \cong 2\times 10^{-3}\frac{H_{END}^2}{M_{pl}},
    \end{eqnarray}
where $g_{reh}=106.75$ are the degrees of freedom in the Standard Model and we have used that 
$\langle \rho_{kin}\rangle \cong   
   2\times 10^{-3} H_{END}^4$.

\

To end this point, choosing $H_{END}\cong 10^{-6} M_{pl}$,  we get a reheating temperature of the order of $T_{reh}\cong 5\times 10^3$ GeV, in agreement with the result obtained in \cite{pv}.

}

\section{Reheating temperature}

The main goal of this
section is to obtain an analytic formula for the reheating temperature via gravitational particle production of heavy massive particles conformally coupled to gravity. To perform the calculation we will use the analytic formula (\ref{rho1}) for the energy density of the produced particles.

\

Then, when the heavy particles decay  into lighter ones -this is a necessary condition in order to arrive at a thermal bath of relativistic particles which will reheat the Universe-  before the end of the kination regime, the reheating temperature as a function of the decay rate $\Gamma$ is given by (see \cite{haro3} for details)

\begin{eqnarray}\label{reheating1}
 T_{\text{reh}}= 
 \left(\frac{30}{\pi^2g_{reh}} \right)^{1/4}
 \langle\rho_{dec}\rangle^{\frac{1}{4}}
 \sqrt{\frac{\langle\rho_{dec}\rangle}{\rho_{\varphi,dec}}}
 = 
  \left(\frac{10}{3\pi^2g_{reh}} \right)^{1/4}
 \left(\frac{\langle\rho_{kin}\rangle^3}{H_{END}^3
 \Gamma M_{pl}^8}\right)^{1/4}M_{pl},
 \end{eqnarray}
where we have assumed that there is no drop of energy in the phase transition,  that is  $H_{kin}\cong  H_{END}$,
the energy density of the inflaton field at the decay is $\rho_{\varphi,dec}=3\Gamma^2M_{pl}^2$, and we have used that at the decay the energy density of the produced particles is related with  its value at the beginning of kination as follows, $\langle \rho_{dec}\rangle
=\langle \rho_{kin}\rangle\frac{\Gamma}{H_{kin}}
\cong\langle \rho_{kin}\rangle\frac{\Gamma}{H_{END}}$, because during  kination the Hubble rate scales as the energy density of the massive produced particles, i.e., as $\left(\frac{a_{kin}}{a(\eta)}  \right)^3$, and the decay ends when $H\cong \Gamma$.

\

In addition, the decay rate satisfies the constraint
\begin{eqnarray}\label{bound}
\frac{\langle\rho_{kin}\rangle}{3H_{END}M_{pl}^2}\leq \Gamma\leq  H_{END},
\end{eqnarray}
which comes from the fact that $\Gamma\leq H_{kin}\cong  H_{END}$ (the decay is after the beginning of kination) and 
$\langle \rho_{dec}\rangle\leq 3\Gamma^2M_{pl}^2$ (the decay is before the end of kination).

\subsection{Maximum reheating temperature}

The maximum reheating temperature, namely $T_{\text{reh}}^{\text{max}}$, is obtained when the decay is produced at the end of kination. This happens because during all the kination period the energy density of the produced particles scales as $a^{-3}$ (recall that after the decay this energy density decays as $a^{-4}$) and for this reason it soon reaches the energy density of the inflaton. Then, choosing 
$
\Gamma=\frac{\langle\rho_{kin}\rangle}{3H_{END}M_{pl}^2}$,  the maximum value of the reheating temperature is

\begin{eqnarray}\label{tmax0}
 T_{\text{reh}}^{\text{max}}(m_{\chi})=  
  \left(\frac{10}{\pi^2g_{reh}} \right)^{1/4}
 \sqrt{\frac{\langle\rho_{kin}\rangle}{H_{END}
  M_{pl}}},
 \end{eqnarray}
 and inserting the value of $\langle\rho_{kin}\rangle$ given by the equation (\ref{rho1})  in this expression, one gets

{
\begin{eqnarray}\label{tmax}
    T_{\text{reh}}^{\text{max}}(m_{\chi})\cong\frac{1}{\pi^2}
    e^{-\frac{\pi m_{\chi}}{4\sqrt{2}H_{END}}}    \left( \frac{5m_{\chi}H_{END}}{16g_{reh}M_{pl}^2} \right)^{1/4}
    m_{\chi}\cong
    2\times 10^{-2}e^{-\frac{\pi m_{\chi}}{4\sqrt{2}H_{END}}}    
    \left( \frac{m_{\chi}H_{END}}{M_{pl}^2} \right)^{1/4}
    m_{\chi},\end{eqnarray}
    where we can see that this maximum reheating temperature depends explicitly on the mass of the produced particles and the value of the Hubble rate at the end of the accelerated period.

\

As a function of  $m_{\chi}$, its maximum value is reached when 
$m_{\chi}=\frac{5\sqrt{2}H_{END}}{\pi}\cong 2.2 H_{END}$, which leads to a reheating temperature of
\begin{eqnarray}
    T_{\text{reh}}^{\text{max}}\cong 2\times 10^{-2}
    \sqrt{\frac{H_{END}}{M_{pl}}}H_{END},
\end{eqnarray}
and choosing $H_{END}\cong 10^{-6} M_{pl}$, which for the majority of models is approximately the value of the Hubble rate at the end of inflation, one obtains
\begin{eqnarray}
T_{\text{reh}}^{\text{max}}\cong 2\times 10^{-11} M_{pl}\cong 5\times 10^7 \mbox{ GeV}.
\end{eqnarray}

\

In addition, for $H_{END}\sim 10^{-6}M_{pl}$ we have calculated:

\

\begin{tabular}{| c | c || c | c |}
\hline
 \hspace{1cm}Mass $(m_{\chi})$ \ \ \ \ \ & Maximum Temperature  &
\hspace{1cm}Mass $(m_{\chi})$ \ \ \ \ \ & Maximum Temperature 
\\ \hline
$5\times 10^{-6} M_{pl}$ & $2\times 10^7$  GeV 
& $10^{-7} M_{pl}$ & $3\times 10^6$  GeV \\
$ 10^{-5} M_{pl}$ & $ 3\times 10^6$  GeV & $10^{-8} M_{pl}$ & 
$  10^5$ GeV
\\
$2 \times 10^{-5} M_{pl}$ & $ 10^4$  GeV & $10^{-10} M_{pl}$ & $5\times 10^2$  GeV\\
$2.5\times 10^{-5} M_{pl}$ & $2\times 10^3$  GeV &$10^{-12} M_{pl}$ &
$1$ GeV
\\
$5\times 10^{-5} M_{pl}$ & $6$  MeV & $10^{-14} M_{pl}$ & $5$ MeV \\\hline
\end{tabular}

\

\

Moreover, since the Nucleosynthesis occurs at temperatures of the order of $1$ MeV, the reheating temperature has to be greater than $1$ MeV in order to ensure its success. Therefore, when the decay is at the end of kination, the viable masses  have to be approximately less than $5\times 10^{-5} M_{pl}$ and greater than $10^{-14} M_{pl}$.

\

}

Finally, dealing with  massive particles non-conformally coupled to gravity, we can use the approximation given in (\ref{master1}) to get the following maximum reheating temperature,
{ \begin{eqnarray}
   T_{\text{reh}}^{\text{max}}(m_{\chi})\cong 2\times 10^{-2} e^{-\pi\frac{ \sqrt{m_{\chi}^2+(1-6\xi)H^2_{END}}  }
   {4\sqrt{2}H_{END}}}
\left(\frac{\sqrt{m_{\chi}^2+(1-6\xi)H_{END}^2}}{H_{END}}\right)^{1/4}
   \sqrt{ \frac{m_{\chi} \sqrt{ m_{\chi}^2+(1-6\xi)H^2_{END} } }{H_{END}M_{pl}}}H_{END}.
\end{eqnarray}}

\subsection{Overproduction of GWs}

The success of the Big Bang Nucleosynthesis (BBN) demands that the ratio of the energy density of the Gravitational Waves (GWs) to the energy density of the produced particles satisfies \cite{hossain}
\begin{eqnarray}\label{constraint}
    \frac{\langle\rho_{GW,reh}\rangle}{\langle\rho_{reh}\rangle}\leq 10^{-2},
\end{eqnarray}
where the energy density of the GWs is {
$\langle\rho_{GW,reh}\rangle\cong 2\times 10^{-3}H_{END}^4\left( \frac{a_{kin}}{a_{reh}}\right)^4$} since they satisfy the same equation as the massless particles minimally coupled to gravity (see eq. (\ref{GW})).

\

In the seminal paper \cite{pv} the authors pointed out the inviability of the reheating via the production of light particles nearly minimally coupled to gravity because they satisfy the same equation as the GWs and, thus, both energy densities scale with the same rate. Then, a way to overcome the constraint 
(\ref{constraint}) is to assume that the reheating is via gravitational production of heavy particles. Effectively, we consider the conformally coupled case and we assume that particles decay at the end of kination, meaning that 
\begin{eqnarray}
 3\Gamma^2M_{pl}^2\cong 3H_{END}^2M_{pl}^2
 \left( \frac{a_{kin}}{a_{reh}}\right)^6\cong 
 \langle \rho_{kin}\rangle \left( \frac{a_{kin}}{a_{reh}}\right)^3,  \end{eqnarray}
and obtaining 
\begin{eqnarray}
\left( \frac{a_{kin}}{a_{reh}}\right)^3\cong
\frac{\langle \rho_{kin}\rangle}{3H_{END}^2M_{pl}^2}, \qquad \mbox{and}\qquad 
\Gamma=\frac{\langle \rho_{kin}\rangle}{3H_{END}M_{pl}^2}.
\end{eqnarray}
Now, from these last results we get {
\begin{eqnarray}
    \frac{\langle\rho_{GW,reh}\rangle}{\langle\rho_{reh}\rangle}\cong 2\times 10^{-3}
    {H_{END}^4}
    \left(\frac{\langle \rho_{kin}\rangle^{-2}}{3H_{END}^2M_{pl}^2}\right)^{1/3}\cong
    6\times 10^{-2}
    \left( \frac{H_{END}}{m_{\chi}}\right)^2 \left( \frac{m_{\chi} H_{END}}{M_{pl}^2}\right)^{1/3}    e^{\frac{\pi m_{\chi}}{3 \sqrt{2}H_{END}}},
\end{eqnarray}}
where we have used our formula (\ref{master1}) with $\xi=1/6$. And, since in the majority of models $H_{END}\cong 10^{-6} M_{pl}$, one arrives at
{\begin{eqnarray}
    \frac{\langle\rho_{GW,reh}\rangle}{\langle\rho_{reh}\rangle}\cong
    6\times 10^{-16}\left( \frac{M_{pl}}{m_{\chi}}\right)^{5/3}    
     \exp\left( {\frac{ 10^6\pi m_{\chi}}{3\sqrt{2}M_{pl}}} \right)   ,\end{eqnarray}
which leads to the constraint 
\begin{eqnarray}
    \left( \frac{M_{pl}}{m_{\chi}}\right)^{5/3}    
     \exp\left( {\frac{ 10^6\pi m_{\chi}}{3\sqrt{2}M_{pl}}} \right)\leq 2\times 10^{13}   ,\end{eqnarray}
which is satisfied  when 
\begin{eqnarray}
   10^{-8}M_{pl}\leq m_{\chi}\leq 1.6\times 10^{-5} M_{pl},
\end{eqnarray}
leading to a maximum reheating temperature bounded by
\begin{eqnarray}
    10^5 \mbox{ GeV}\lesssim  T_{\text{reh}}^{\text{max}}
    \lesssim 5\times 10^7 \mbox{ GeV}.    \end{eqnarray}

}

\subsection{Dark matter via gravitational particle production}

Here we consider two kind of particles, on the one hand the $X$-particles which will decay into lighter ones to reheat the Universe and on the other hand the $Y$-particles which represent the dark matter. 

\

To simplify, we will assume that they are conformally coupled to gravity and, thus, its energy density at the beginning of kination is given by the formula (\ref{master1}) with $\xi=1/6$. As a consequence, at the reheating time we have {
\begin{eqnarray}
    \langle \rho_{Y,reh}\rangle=
    e^{-\frac{\pi(m_Y-m_X)}{2\sqrt{2}H_{END}}}\left( \frac{m_Y}{m_X}\right)^{5/2}
    \langle \rho_{X,reh}\rangle.
    \end{eqnarray}
    }

\

After the reheating the energy density of the $X$-particles scale as $a^{-4}$. So, at the matter-radiation equality  we have {
\begin{eqnarray}
  \frac{a_{reh}}{a_{eq}}=\frac{\langle \rho_{Y,reh}\rangle}{\langle \rho_{X,reh}\rangle}=e^{-\frac{\pi(m_Y-m_X)}{2\sqrt{2}H_{END}}}\left( \frac{m_Y}{m_X}\right)^{5/2} 
\end{eqnarray}}
and, hence, {
\begin{eqnarray}
\langle \rho_{Y,eq}\rangle=
\langle \rho_{Y,reh}\rangle
\left( \frac{a_{reh}}{a_{eq}}\right)^3=e^{-\frac{\sqrt{2}\pi(m_Y-m_X)}{H_{END}}}\left( \frac{m_Y}{m_X}\right)^{10}\frac{\pi^2 g_{reh}}{30}T^4_{rh}(m_X),\end{eqnarray}}
where as a reheating temperature we will choose the maximum one, i.e. $T_{rh}(m_X)=\left( \frac{10}{\pi^2 g_{reh}}\right)^{1/4}\sqrt{
\frac{\langle \rho_{X,kin}\rangle}{H_{END}M_{pl}}}$.

\

Then, after some algebra we get {
\begin{eqnarray}
\langle \rho_{Y,eq}\rangle=
\frac{1}{96\pi^4}e^{-\frac{\pi(2m_Y-m_X)}{ \sqrt{2}H_{END}}}
\frac{m_Y^{10}H_{END}}{m_X^5M_{pl}^2}.
\end{eqnarray}}

\

On the other hand, from the observational data we know that
(see for instance \cite{haro4})
\begin{eqnarray}
\langle \rho_{Y,eq}\rangle\cong
3\times 10^{-121}(1+z_{eq})^3 M_{pl}^4,
\end{eqnarray}
where the red-shift at the matter-radiation equality can be approximated by $z_{eq}\cong 3365$.

\

After equating both expressions and taking $H_{END}=10^{-6}M_{pl}$, 
we obtain the following relation between both masses, {
\begin{eqnarray}
 \exp\left(-\frac{\pi(2m_Y-m_X)10^5}{\sqrt{2}M_{pl}}  \right) \cong  10^{-10}\frac{\sqrt{M_{pl}m_X}}{m_Y},  
\end{eqnarray}}
which is equivalent to {
\begin{eqnarray}
    2\bar{Y}-\bar{X}\cong \frac{15}{2}\cong 17.27,
\end{eqnarray}}
where we have introduced the notation {
\begin{eqnarray}
\bar{A}=\frac{\pi}{\sqrt{2}}
  \frac{10^5m_A}{M_{pl}}-\frac{1}{2}\ln\left(\frac{10^5m_A}{M_{pl}}\right),  \end{eqnarray}}
with $A=X,Y$.

\

Finally, since the viable masses of the $X$-particles -which overcome the problem of the overproduction of GWs- satisfy  {
$ 10^{-8}M_{pl}\leq m_{\chi}\leq 
    1.6\times 10^{-5} M_{pl}$,}
one concludes that,
if the dark matter is produced gravitationally, then the mass of the dark energy must approximately belong to the domain {
\begin{eqnarray} 
10^{-14} \leq\frac{m_{\chi}}{M_{pl}} \leq 4\times 10^{-14}  \qquad \mbox{and}\qquad  
 4.7\times 10^{-5} \leq\frac{m_{\chi}}{M_{pl}} \leq 5 \times 10^{-5} .
\end{eqnarray}}

\subsection{A toy model}\label{toy-model}

\begin{figure}[H]
    \centering
    \includegraphics[width=200pt]{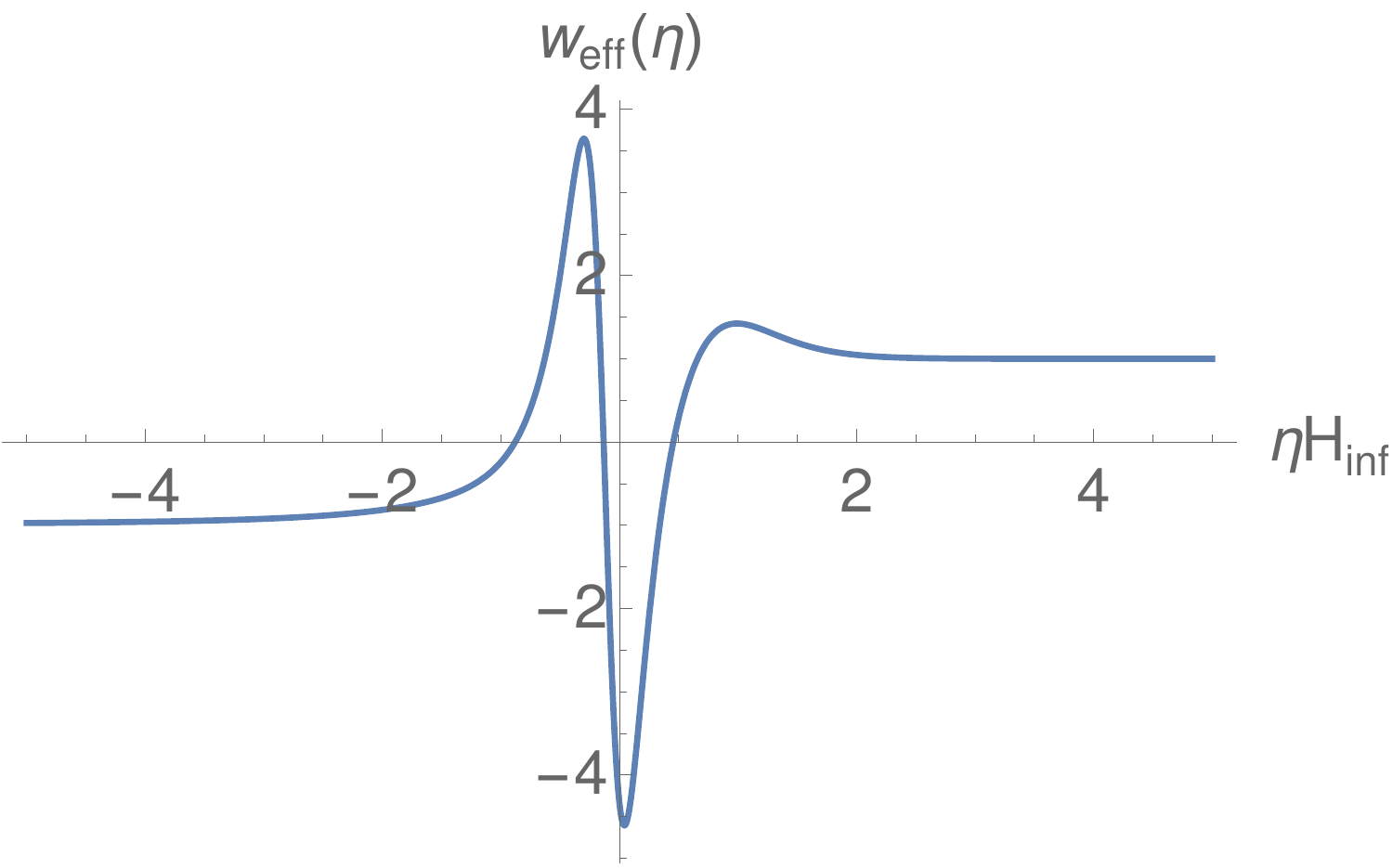}
    \caption{EoS parameter for the model (\ref{background1}) when $\Delta\eta=\frac{0.3}{H_{inf}}$ and $H_{inf}=10^{-6} M_{pl}$.}
    \label{fig:my_label2}
\end{figure}

In \cite{hashiba} the authors introduce 
the following  smooth scale factor
\cite{hashiba},
\begin{eqnarray}\label{background1}
a^2(\eta)=\frac{1}{2}\left[
\left(1-\tanh(\eta/\Delta\eta)\right)\frac{1}{1+H_{inf}^2\eta^2}+
(1+\tanh(\eta/\Delta\eta))(1+H_{inf}\eta)
\right],   
\end{eqnarray}
which contains a 
 super-kination ($w_{eff}>1$) phase during the phase transition from inflation to kination as we can see in Figure \ref{fig:my_label2}. This is due to the fact that the corresponding potential is negative during the phase transition, this unusual fact only happens in a few models of Quintessential Inflation, for example in Lorentzian Quintessential Inflation \cite{benisty}. 
However, it also has a phantom phase ($w_{eff}<-1$) which is produced by a phantom field whose energy density is  $\rho=-\frac{\dot{\varphi}^2}{2}+V(\varphi)$, which never happens in Quintessential Inflation where the dynamics is driven by  a non-phantom scalar field.

\

In addition,  during the phase transition from the end of the slow-roll to the beginning of kination there are three moments where $w_{eff}=-1/3$, so it is not clear at all when the early accelerated expansion finishes.

\

For this model with $\Delta\eta=\frac{0.3}{H_{inf}}$ and $H_{inf}=10^{-6} M_{pl}$,  in \cite{hashiba} the authors obtained {\it empirically} in order to match with their numerical results that the energy density of the massive particles conformally coupled to gravity is {
\begin{eqnarray}\label{has}
   \langle \rho_{kin}\rangle\cong  
   \frac{A^4g_{reh}\pi^2}{30}\left(\frac{m_{\chi}}{H_{inf}}\right)^{4d}
   e^{-4m_{\chi}\Delta\eta} H_{inf}^4 {a_{kin}^{-3}}\cong 
    3\times 10^{-4}e^{-4m_{\chi}\Delta\eta}
    \sqrt{\frac{m_{\chi}}{H_{inf}}}H_{inf}^2m_{\chi}^2,
   \end{eqnarray}
where  $A=0.052$ and $d=0.62$, and we have also made the  approximations
$a_{kin}\cong 1$ and $4d=2.48\cong 2.5$.}

\

\begin{figure}[H]
    \centering
    \includegraphics[width=230pt]{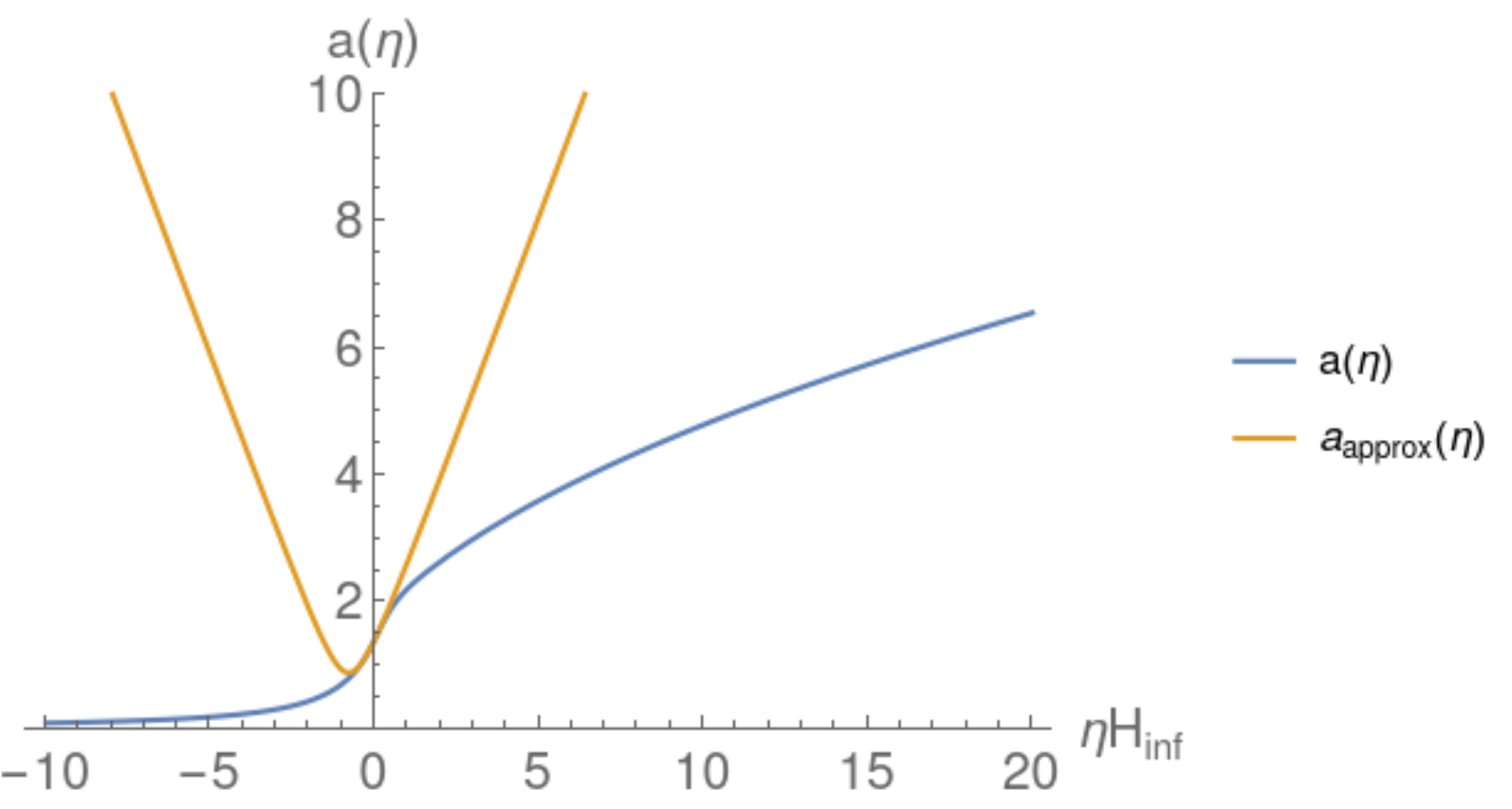}
    \includegraphics[width=230pt]{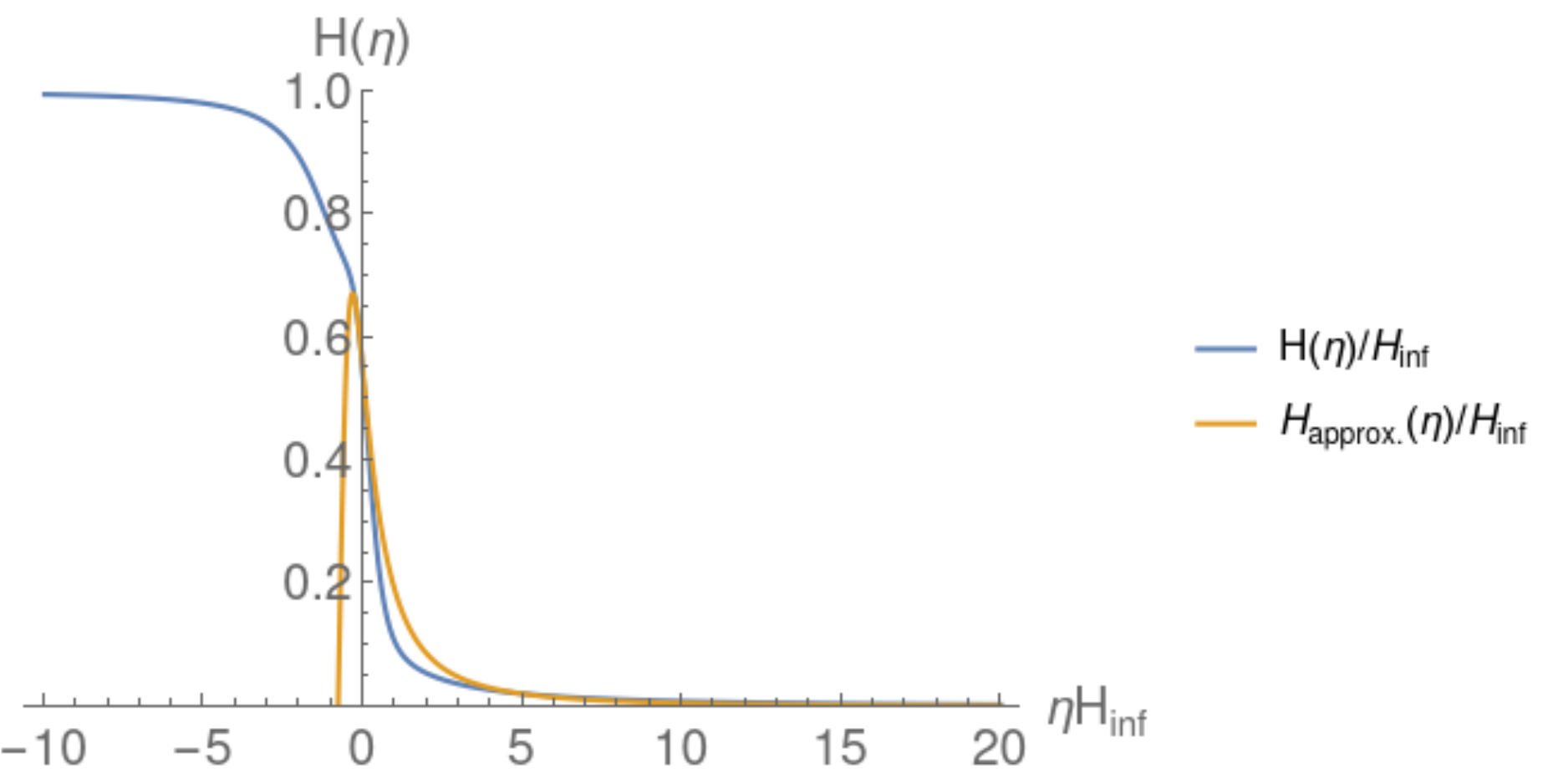}
    \includegraphics[width=200pt]{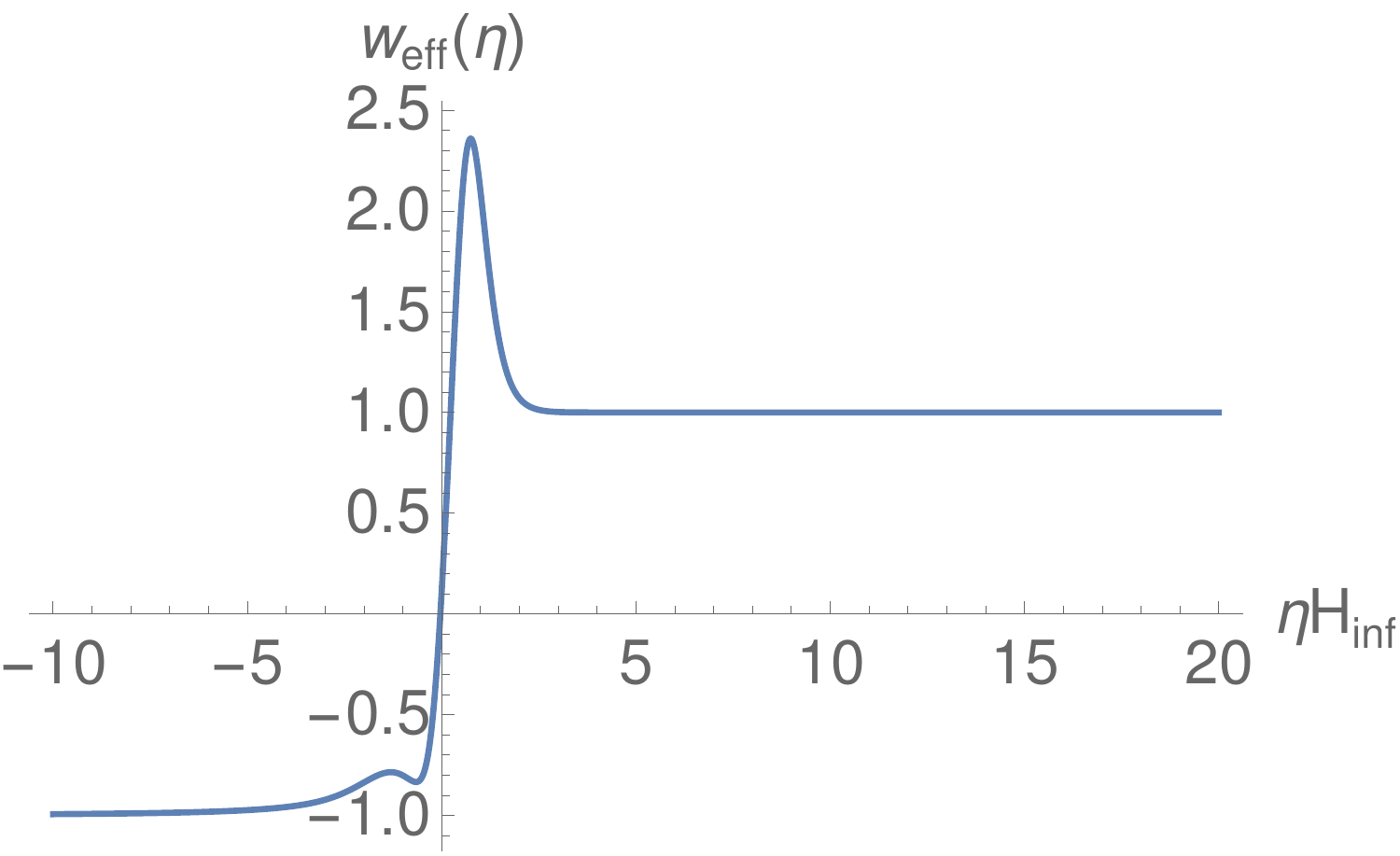}
    \caption{Pictures of the scale factor, Hubble rate and EoS parameter for the model (\ref{background}) for $H_{inf}=10^{-6} M_{pl}$ and $\Delta\eta=\frac{1}{2H_{inf}}$. We can see that at early times the universe is in an inflationary phase ($w_{eff}=-1$) and at late times in a kination one ($w_{eff}=1$).}
    \label{fig:my_label1}
\end{figure}

Here, in order to remove the phantom phase, we improve the model 
 (\ref{background1}) as follows (see Figure  \ref{fig:my_label1}):
\begin{eqnarray}\label{background}
a^2(\eta)=\frac{1}{2}\left[
\left(1-\tanh(\eta/\Delta\eta)\right)\frac{1}{1+H_{inf}^2\eta^2}+
(1+\tanh(\eta/\Delta\eta))(3+2H_{inf}\eta)
\right].    
\end{eqnarray}

\begin{figure}[H]
    \centering
    \includegraphics[width=230pt]{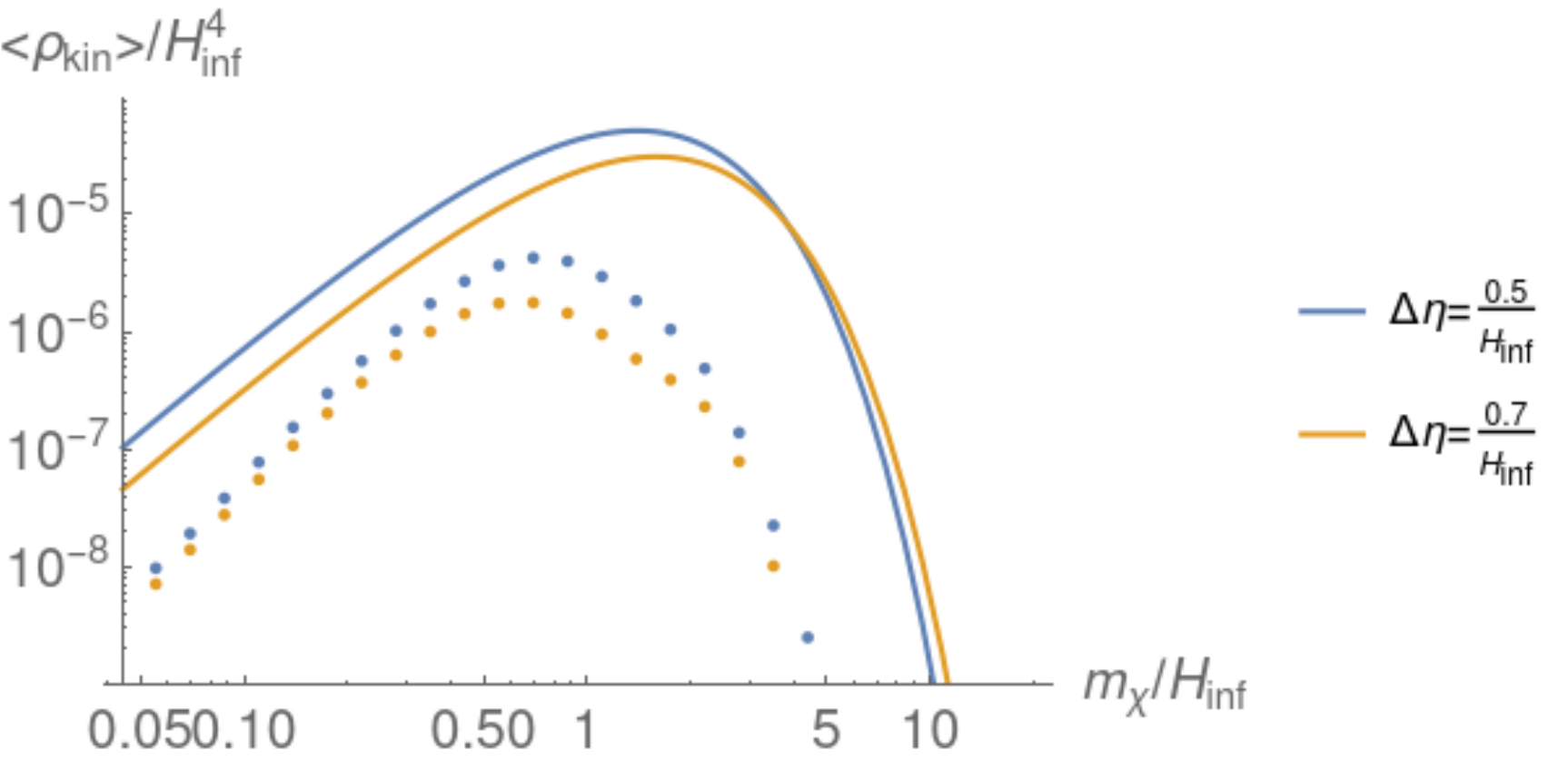}\hspace{1cm}
    \includegraphics[width=190pt]{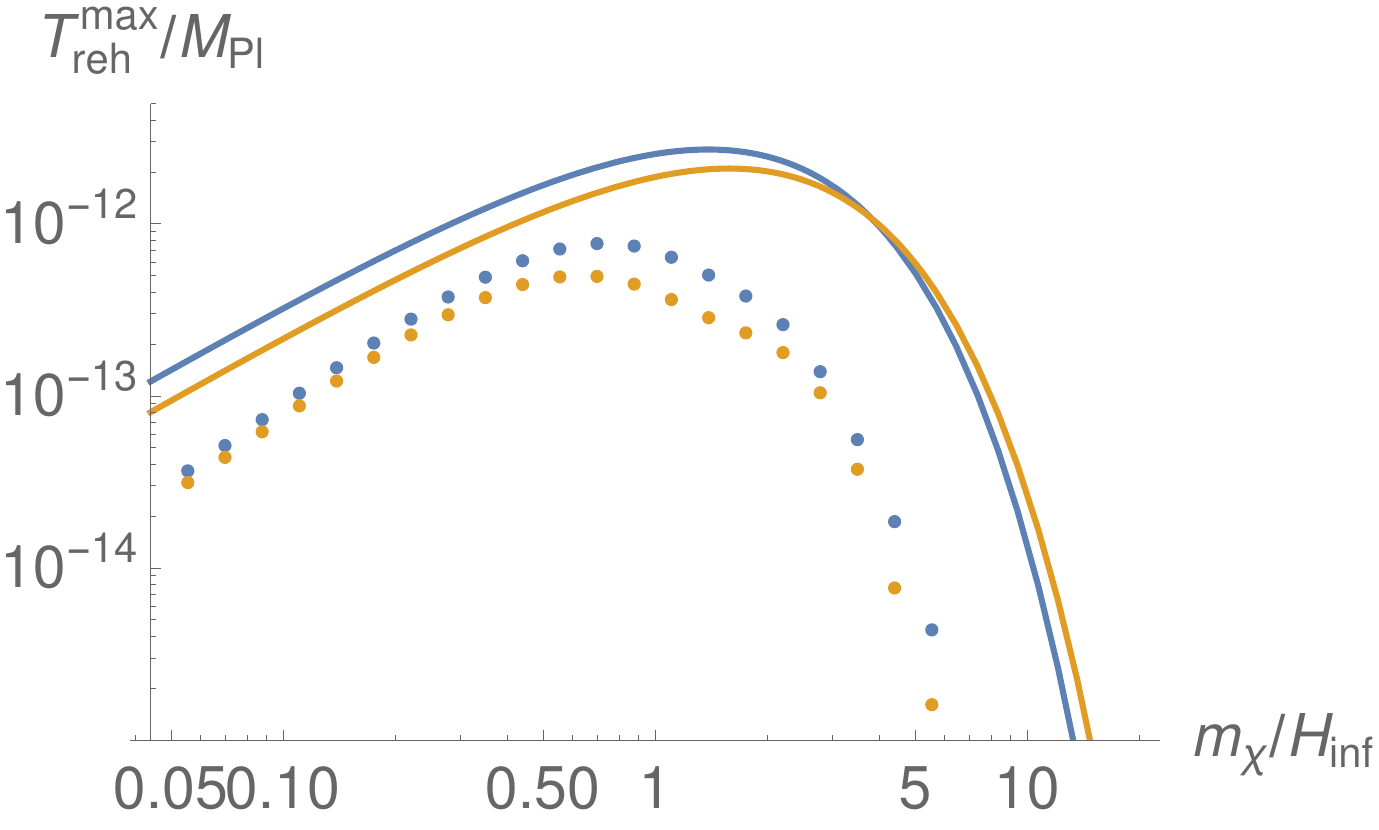}
    \caption{Analytic and numerical values of the energy density and maximum reheating temperature of the model \eqref{background} for $\Delta\eta=\tfrac{0.5}{H_{inf}}$ and $\Delta\eta=\tfrac{0.7}{H_{inf}}$. In dots the numerical results}
    \label{fig:rhotemps}
\end{figure}

{For our model \eqref{background} the phase transition, where particles are created, has an approximate duration of $4\Delta\eta$, because $\tanh (\pm 2)\cong \pm 1$. Then,   for $\Delta\eta=\frac{1}{2H_{inf}}$ it means that the phase transition occurs in the period $[-\frac{1}{H_{inf}}, \frac{1}{H_{inf}}]$, where our approximation matches very well, but it starts to deviate after the beginning of kination, and as we have already explained  our approximation will be more accurate when we better approximate the frequency (in this case the scale factor) during the phase transition and a small enough period after kination.

\

For our model  the energy densities and the corresponding maximum temperatures are depicted in Figure \ref{fig:rhotemps}, where we can see that   our maximum reheating temperatures  obtained from the analytic approximation 
    only differ in less than an order to
     the results obtained numerically for the masses that lead a maximum reheating temperature greater that $5\times 10^{-14}M_{pl}\cong   10^{5}$ GeV. For greater masses the approximation is worse,  but it does not matter too much, because   for these masses the reheating temperature is too small due to the exponential decrease as a function of the mass.}

\

{
Coming back to the empirical  formula (\ref{has}) obtained in \cite{hashiba}, when we apply it to our model (\ref{background}) with $\Delta\eta=\frac{1}{2H_{inf}}$ with $H_{inf}=10^{-6} M_{pl}$ we have checked numerically that $H_{END}=6\times 10^{-7} M_{pl}$, thus the formula (\ref{has}) becomes (see the equation (3.9) of \cite{hashiba} where the authors explain that they have disregarded the square root)
\begin{eqnarray}\label{has1}
   \langle \rho_{kin}\rangle\cong   
    6\times 10^{-4}e^{-1.2\frac{m_{\chi}}{H_{END}} }
    \sqrt{\frac{m_{\chi}}{H_{END}}}H_{END}^2m_{\chi}^2,
   \end{eqnarray}
which has to be compared with our formula (\ref{rho1}), that is, with 
\begin{eqnarray}\label{haro}
   \langle \rho_{kin}\rangle\cong   
    7\times 10^{-3}e^{-1.11\frac{m_{\chi}}{H_{END}} }
    \sqrt{\frac{m_{\chi}}{H_{END}}}H_{END}^2m_{\chi}^2.
   \end{eqnarray}
We can see that they have practically the same shape and differ in one order. In addition since the maximum reheating temperature is proportional to the square root of the energy density at the kination (see (\ref{tmax0})), one can deduce that the corresponding maximum temperatures differ less than an order.
In fact, applying the formula (\ref{tmax}) to (\ref{has1}) one gets
\begin{eqnarray}
    T^{\text{max}}_{\text{reh}}\cong 4\times 10^{-2}
    e^{-0.6\frac{m_{\chi}}{H_{END}} }
    \left( \frac{m_{\chi}H_{END}}{M_{pl}^2} \right)^{1/4} m_{\chi},
    \end{eqnarray}
and applying (\ref{tmax}) to (\ref{haro}) one obtains
\begin{eqnarray}
    T^{\text{max}}_{\text{reh}}\cong 10^{-1}
    e^{-0.55\frac{m_{\chi}}{H_{END}} }
    \left( \frac{m_{\chi}H_{END}}{M_{pl}^2} \right)^{1/4} m_{\chi}.
    \end{eqnarray}

}

\begin{remark}
Dealing with instant preheating where, for conformally coupled particles, the frequency is given by \cite{fkl0,fkl}
\begin{eqnarray}\label{instant0}\omega_k^2(\eta)=k^2+m_{\chi}^2a^2(\eta)+g^2a^2(\eta)
(\varphi(\eta)-\varphi_{kin})^2,\end{eqnarray}
being $g$ a dimensionless interaction parameter between the inflation field $\varphi$ and the quantum field,  
the analytic computation of the Bogoliubov coefficients is based on the approximation
$\varphi(\eta)-\varphi_{kin}\cong \varphi_{kin}'(\eta-\eta_{kin})$  and the assumption that  the  universe is static with $a(\eta)=a_{kin}$. Then, the frequency becomes 
\begin{eqnarray}\label{instant1}
\omega_k^2(\eta)=k^2+m_{\chi}^2a^2_{kin}+g^2a^2_{kin} (\varphi'_{kin})^2(\eta-\eta_{kin})^2,\end{eqnarray}
{and in Figure \ref{fig:w-instant} we can see  in blue (resp. in orange) the plot of (\ref{instant0}) (resp. (\ref{instant1})) for $m_{\chi}=H_{inf}$ and $k=0$, where we can check that the approximation is very good at the phase transition, which in instant preheating occurs at the beginning of kination (for the model \eqref{background} approximately at $\eta_{kin}\cong \frac{5}{2H_{kin}}$), and during a small period of time after it.
\begin{figure}[H]
   \centering
   \includegraphics[width=200pt]{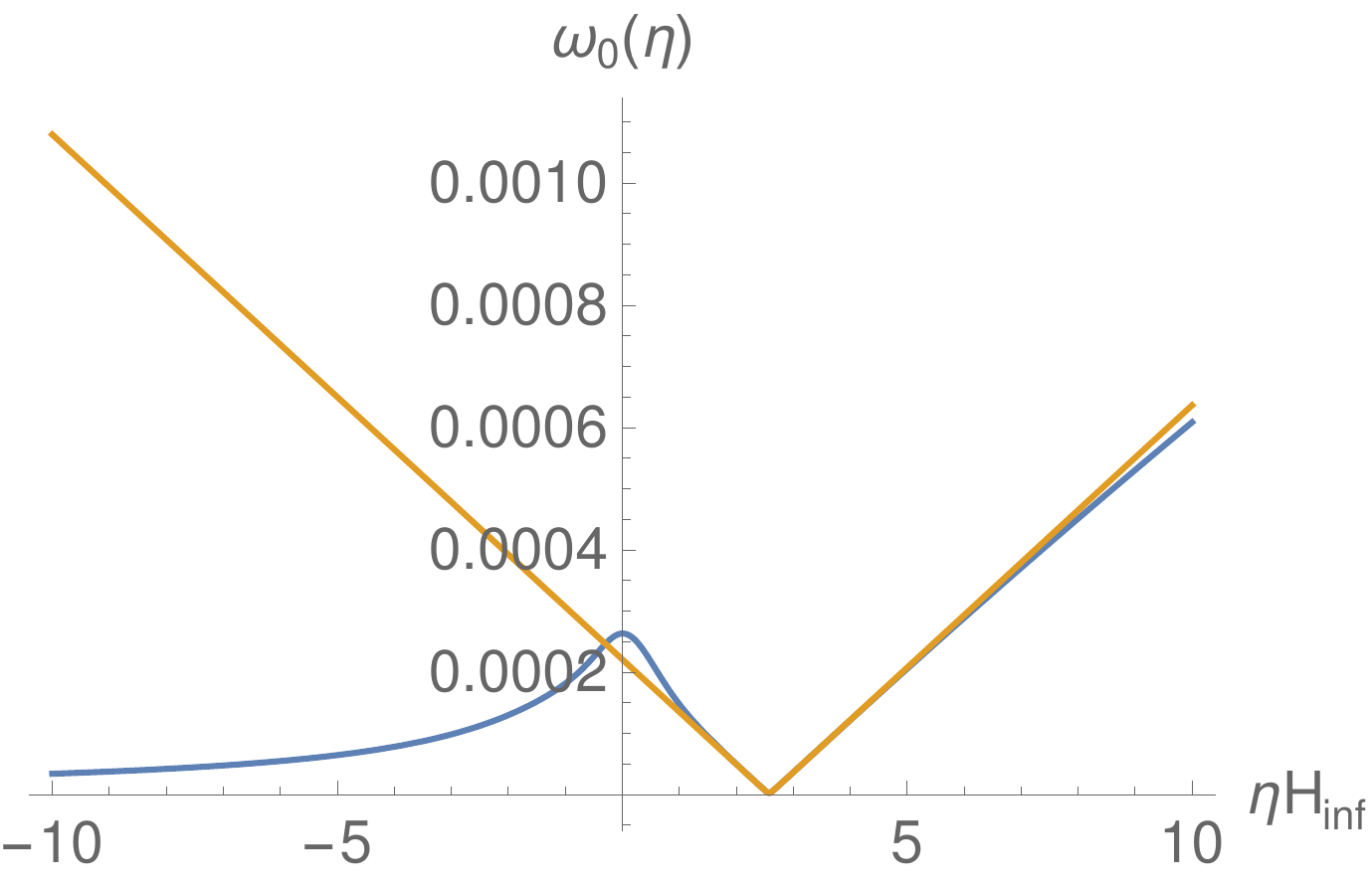}
   \caption{For $g=10^{-4}$, $\Delta\eta=\frac{1}{2H_{inf}}$, $H_{inf}=10^{-6} M_{pl}$, $m_{\chi}=H_{inf}$ and $k=0$, the analytic and numerical plots of  the frequency. }
   \label{fig:w-instant}
\end{figure}}

\

In this case the analytic value of the $\beta$-Bogoliubov coefficients is given by
\begin{eqnarray}\label{instant}
    |\beta_k|^2\cong \exp \left({-\frac{\pi (k^2+m_{\chi}^2a^2_{kin})}{ga_{kin}\varphi'_{kin}}}\right)
    = \exp\left({-\frac{\pi (k^2+m_{\chi}^2a^2_{kin})}{\sqrt{6}ga_{kin}^2H_{kin}M_{pl}}}\right)\cong 
\exp\left({-\frac{\pi (k^2+m_{\chi}^2a^2_{END})}{\sqrt{6}ga_{END}^2H_{END}M_{pl}}}\right),
   \end{eqnarray}
which is approximately   our formula (\ref{beta2}) replacing $\sqrt{3}gM_{pl}$
by $m_{\chi}$.

\

This last formula was tested numerically in \cite{campos} for the original Peebles-Vilenkin model \cite{pv}. Here,
to check this analytic formula we  deal with the  scale factor (\ref{background}), because it does not contain any phantom phase and thus, the inflaton field (also named {\it cosmon} in Quintessential Inflation \cite{rubio}) can be calculated from the
equation 
\begin{eqnarray}
    {\mathcal H}'-{\mathcal H}^2=-
    \frac{1}{2M_{pl}^2}(\varphi')^2,
\end{eqnarray}
where ${\mathcal H}=a'/a$ denotes the conformal Hubble rate.

\

In Figure (\ref{fig:instant}) we have computed the energy density of the produced particles, the numerical calculation has also been  done assuming that the quantum field is in the vacuum only few e-folds before the end of inflation,   and we see  that the approximation of the Bogoliubov coefficient (\ref{instant}) matches very well with the exact numerical calculation using the frequency (\ref{instant0}).

\begin{figure}[H]
   \centering
   \includegraphics[width=190pt]{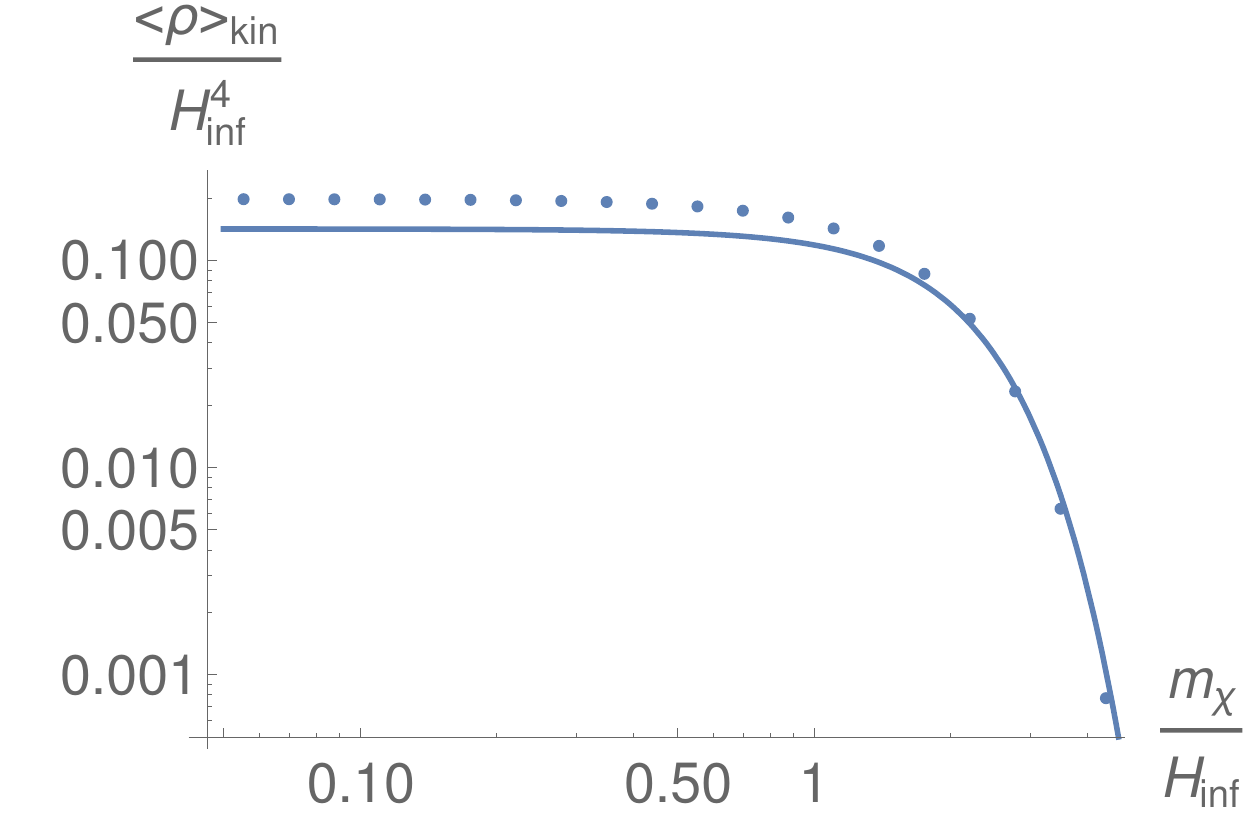}
   \caption{For $g=10^{-4}$, $\Delta\eta=\frac{1}{2H_{inf}}$ and $H_{inf}=10^{-6} M_{pl}$, the analytic and numerical plots of the energy density at the beginning of kination. }
   \label{fig:instant}
\end{figure}

\

Finally, our choice of $g=10^{-4}$ is not random. In inflationary models, so that the vacuum polarization effects do not disturb the last stages of inflation, it is mandatory that $g\gg 10^{-6}$
(see \cite{fkl}). On the other hand, 
to prevent that relic products such as gravitinos or modulus fields -which could appear in supergravity or superstring theories- affect the BBN success, the reheating temperature has to be less than $10^9$ GeV (see for instance \cite{ellis}). Then,  we have the constraint
\begin{eqnarray}
   T^{\text{min}}_{\text{reh}}(m_{\chi})\leq 10^9 \mbox{GeV}. 
\end{eqnarray}

\

For simplicity we consider the case where the bare mass $m_{\chi}$ vanishes. Then, the energy density at the kination is given by
$\langle \rho_{kin}\rangle=\frac{g^2\dot{\varphi}^2_{kin}}{4\pi^4}=
\frac{3g^2H_{kin}^2M_{pl}^2}{4\pi^4}\cong \frac{3g^2H_{END}^2M_{pl}^2}{4\pi^4}$, where once again we assume that there is no drop of energy during the phase transition. Since the minimum reheating temperature is given by
\begin{eqnarray}
    T^{\text{min}}_{\text{reh}}(0)=\left(\frac{10}{3\pi^2g_{reh}} \right)^{1/4}
    \frac{\langle \rho_{kin}\rangle^{3/4}}{H_{END}M_{pl}}\cong 6\times 10^{-3}g^{3/2}\sqrt{H_{END}M_{pl}},
    \end{eqnarray}
and taking into account that for this model $H_{END}\cong 6\times 10^{-7} M_{pl}$, one can conclude that $g\leq 2\times 10^{-3}$, that is, the parameter $g$ is constrained as follows,
\begin{eqnarray}
    10^{-6}\ll g<2\times 10^{-3}.
\end{eqnarray}

\end{remark}

\section{$\alpha$-attractors}

\begin{figure}[H]
   \centering
   \includegraphics[width=200pt]{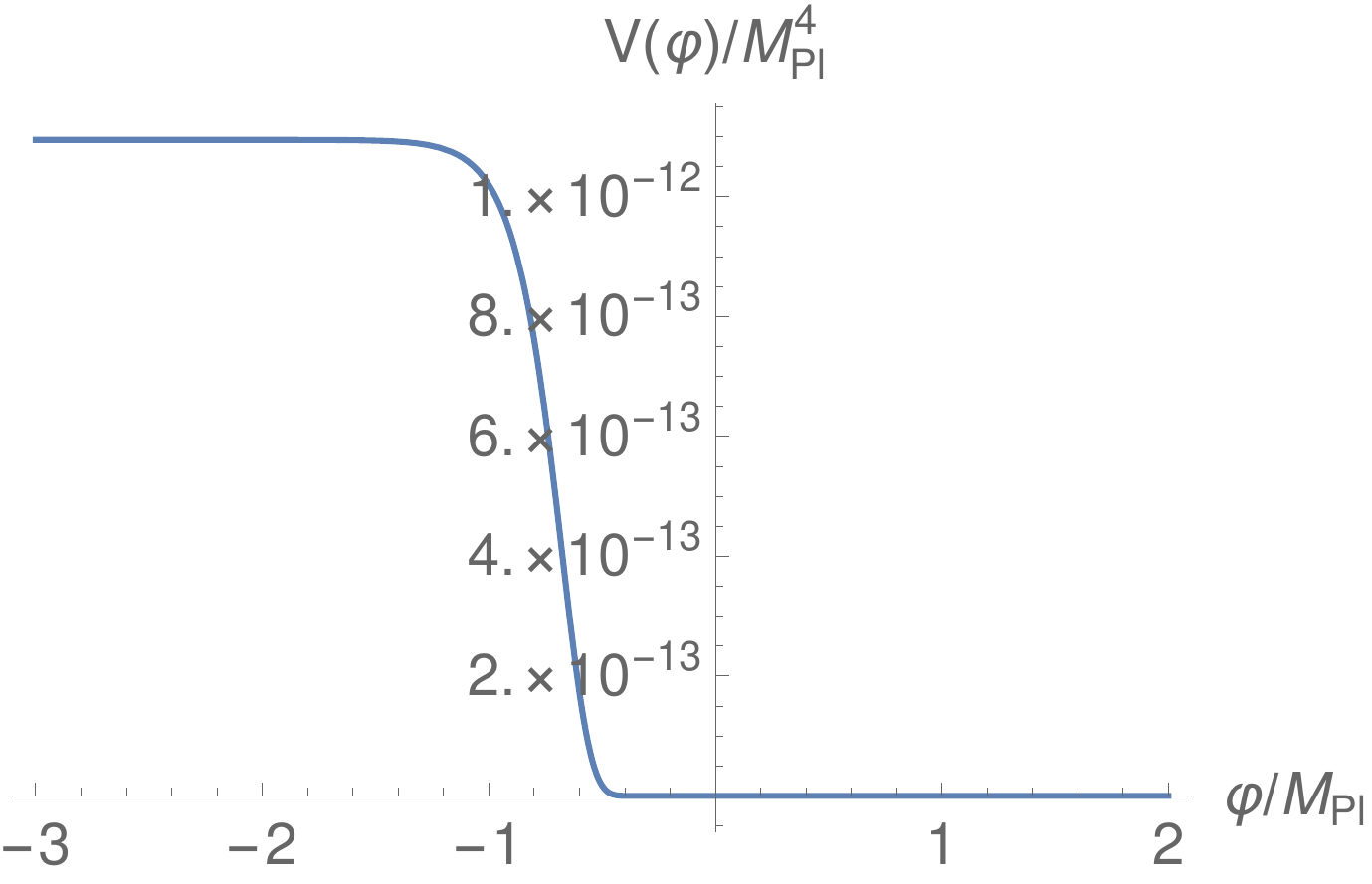}
   \includegraphics[width=200pt]{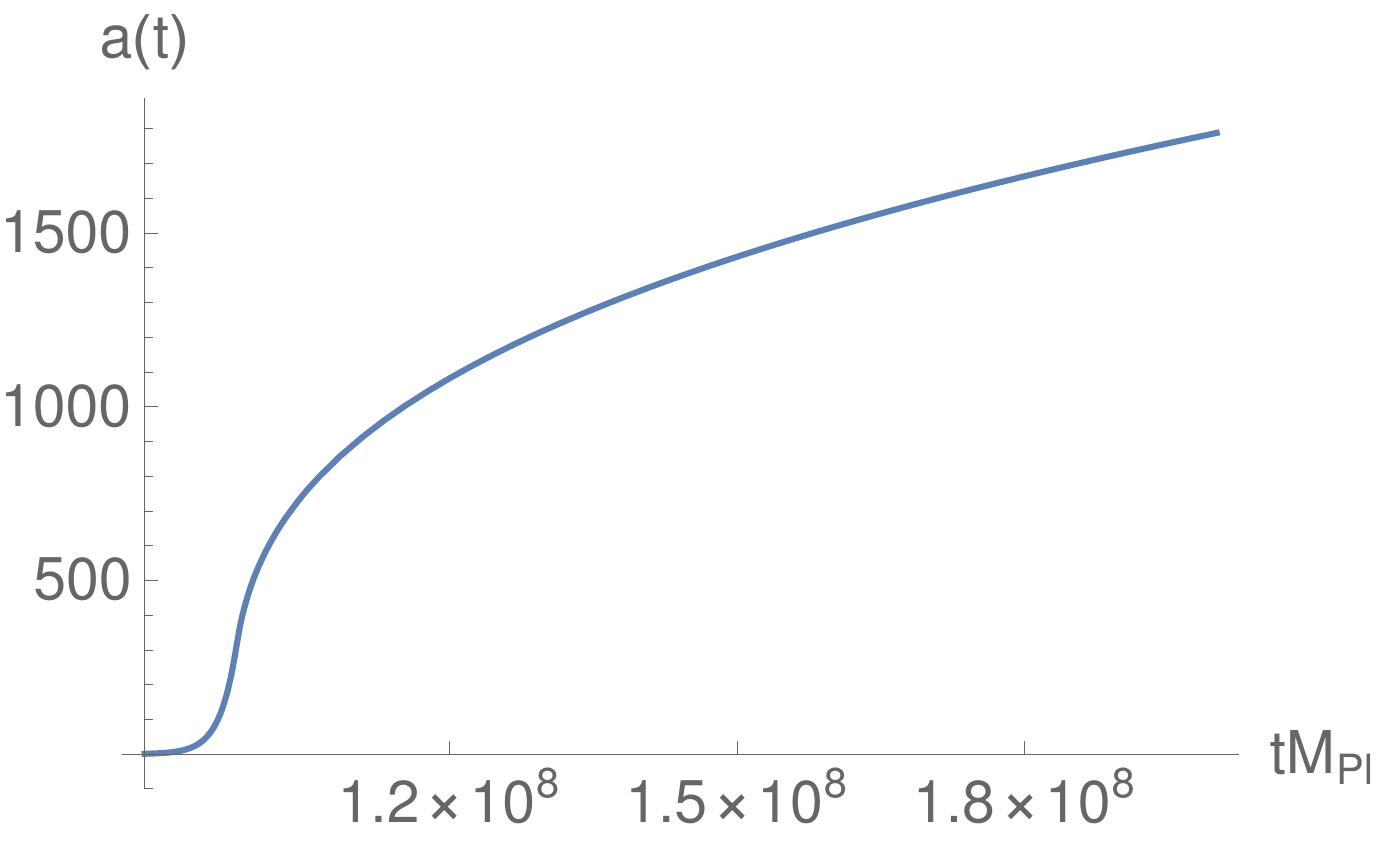}
    \includegraphics[width=200pt]{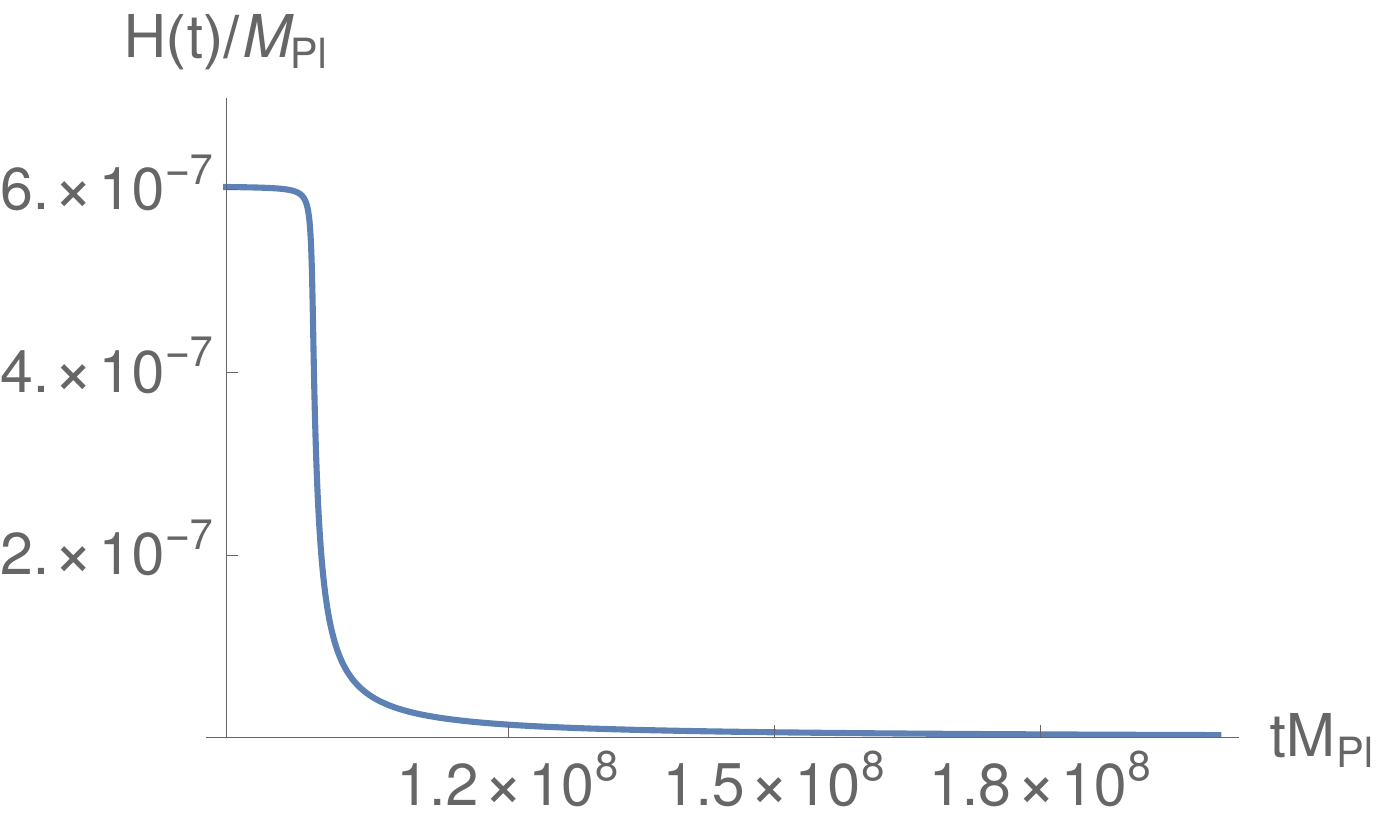}
   \includegraphics[width=200pt]{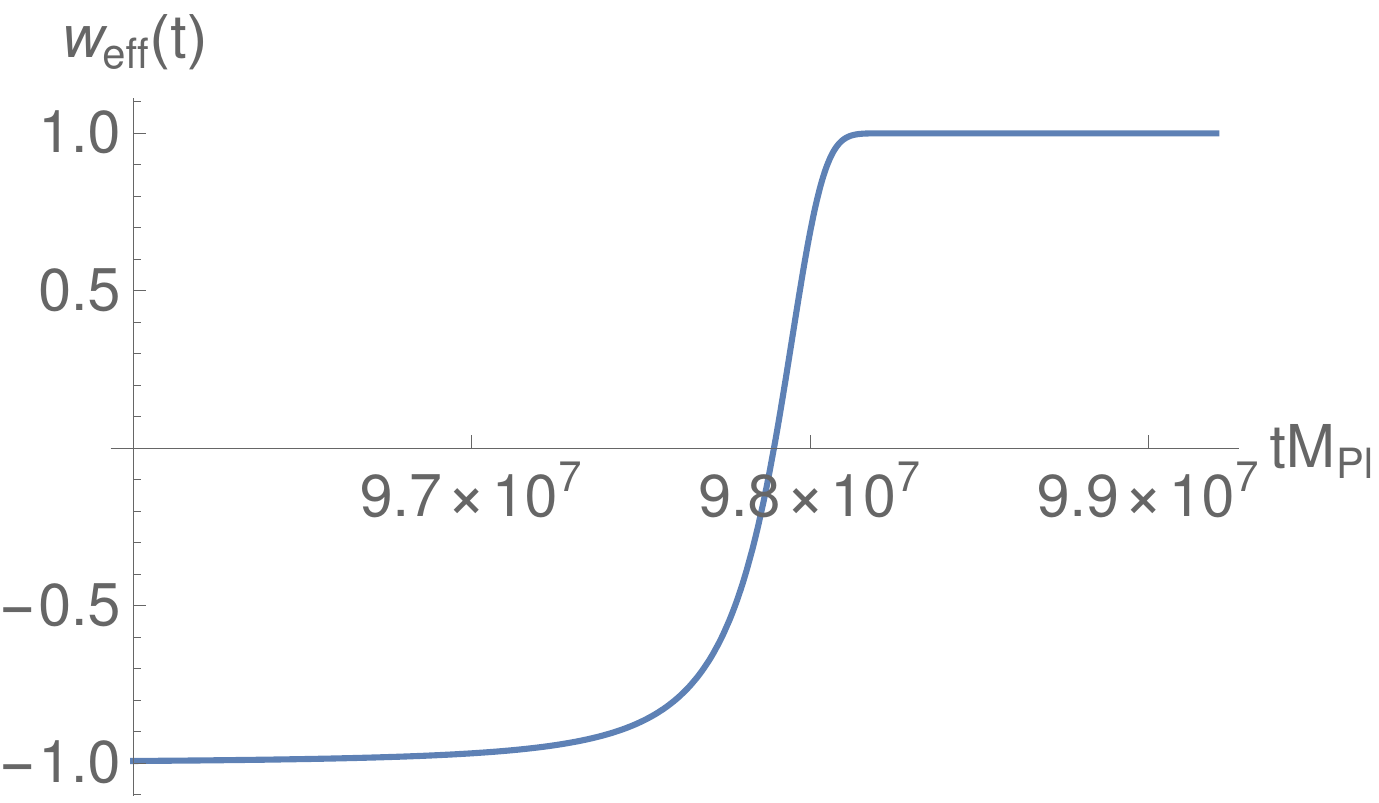}
   \caption{ The exponential $\alpha$-attractor potential, scale factor, Hubble constant and EoS parameter for $\alpha=10^{-2}$ and $n=124$.}
   \label{fig:pot}
\end{figure}

We consider the following  exponential $\alpha$-attractor potential, 
displayed in Figure \ref{fig:pot} \cite{haro},
\begin{eqnarray}\label{alpha}
V(\varphi)=\lambda M_{pl}^4e^{-n\tanh\left(\frac{\varphi}{\sqrt{6\alpha}M_{pl}} \right)},
\end{eqnarray}
where $\lambda$, $\alpha$ and $n$ are dimensionless parameters which have to satisfy the following relations in order to match with the current observation data,
\begin{eqnarray}\label{parameters}
\frac{\lambda}{\alpha}e^{n}\sim 10^{-10} \qquad \mbox{and}\qquad \lambda e^{-n}\sim 10^{-120}.
\end{eqnarray}
  
\

To find analytically $H_{END}$ we calculate the slow-roll parameter
\begin{eqnarray}
    \epsilon=\frac{M_{pl}^2}{2}\left( \frac{V_{\varphi}}{V} \right)^2=\frac{n^2}
    {12\alpha}\frac{1}{\cosh^4\left(\frac{\varphi}{\sqrt{6\alpha}M_{pl}} \right)}.
\end{eqnarray}

Since inflation ends when $\epsilon=1$
and
noting  that
$
\mathop{\mathrm{arccosh}}(x)=\ln(x-\sqrt{x^2-1}),
$
one has
\begin{eqnarray}
    \varphi_{END}=\sqrt{6\alpha}\ln\left( \frac{\sqrt{n}}{(12\alpha)^{1/4}}-\sqrt{\frac{n}{\sqrt{12\alpha}}-1}
    \right) M_{pl}.
\end{eqnarray}

Then, from equation \eqref{parameters} one obtains
\begin{eqnarray}
V(\varphi_{END})=\lambda M_{pl}^4e^{n\sqrt{1-\frac{\sqrt{12\alpha}}{n}}}\cong \lambda M_{pl}^4e^{n\left(1-\frac{\sqrt{3\alpha}}{n}\right)}\cong \alpha e^{-\sqrt{3\alpha}}10^{-10} M_{pl},
\end{eqnarray}
and using that $\rho_{END}=\frac{3V(\varphi_{END})}{2}$ one gets
\begin{eqnarray}
    H_{END}\cong \sqrt{\frac{\alpha}{2}}e^{-\sqrt{3\alpha}/2} 10^{-5} M_{pl}\cong 
    \sqrt{\frac{\alpha}{2}} 10^{-5} M_{pl},\end{eqnarray}
    meaning that  from (\ref{tmax}) the maximum reheating temperature in the conformally coupled case   is given by {
\begin{eqnarray}
T_{\text{reh}}^{\text{max}}(m_{\chi})\cong 6\times 10^{-4} 
\left(\frac{\alpha}{2}\right)^{1/8}\exp\left({-\frac{ \pi \times 10^5  m_{\chi}}{4\sqrt{\alpha}M_{pl}}}\right) 
\left(\frac{m_{\chi}}{M_{pl}} \right)^{1/4} m_{\chi},
\end{eqnarray}
    which for $\alpha=10^{-2}$ becomes
\begin{eqnarray}
T_{\text{reh}}^{\text{max}}(m_{\chi})\cong 3\times 10^{-4} 
\exp\left(-\frac{ \pi\times 10^6  m_{\chi}}{4M_{pl}} \right)
\left(\frac{m_{\chi}}{M_{pl}} \right)^{1/4}
m_{\chi}.
\end{eqnarray}}

\subsection{Numerical calculations}

To contrast this theoretical result with the numerics we have 
to obtain numerically  the background. To do it    one has to  integrate  the conservation equation for the inflaton field, which  in terms of the cosmic time is 
\begin{eqnarray}\label{conservation}
\ddot{\varphi}+3H\dot{\varphi}+V_{\varphi}=0,
\end{eqnarray}
where $H=\frac{1}{\sqrt{3}M_{pl}}\sqrt{\frac{\dot{\varphi}^2}{2}+V(\varphi)  }$. We can choose the initial conditions at the horizon crossing,
 i.e., when the pivot scales leaves the Hubble radius, because at that moment
the system is in the slow-roll phase and, since this regime is an attractor, one has to take initial conditions in the basin of attraction of the slow-roll solution. Then, we take
$\varphi=\varphi_*$ and $\dot{\varphi}=-\frac{V_{\varphi}(\varphi_*)}{3H_*}$, where the ``star" denotes that the quantities are evaluated
at the horizon crossing.

\

Once one has obtained the evolution of the background and in particular the evolution of the Hubble rate, we compute the evolution of the scale factor, which is 
given by 
\begin{eqnarray}
a(t)=a_*e^{\int_{t_*}^t H(s)ds},
\end{eqnarray}
where the value of $a_*$ is arbitrary and can be chosen to be $a_*=1$.

\

Dealing with conformally coupled particles, the Bogoliubov coefficients satisfy the dynamical system (\ref{Bogoliubovequation1}), where one has to replace $\Omega_k$ by $\omega_k$. 
The way to calculate the value of $\beta_k$ (the value of the $\beta$-Bogoliubov coefficient when it stabilizes) is to solve numerically the equation (\ref{kg1}), with initial conditions at late time, for example,  at the horizon crossing, given by  
\begin{eqnarray}
 \chi_{k}(\eta_*)=
 \frac{1}{\sqrt{2\omega_k(\eta_*)}} , \quad
 \chi_{ k}'(\eta_*)=
-i \omega_k(\eta_*)\chi_{ k}(\eta_*). \end{eqnarray}

This means that at that moment the quantum field is at the vacuum. In fact, for the relevant modes the quantum field is in the adiabatic vacuum at the horizon crossing and it does not matter if one chooses  an earlier time  for the initial conditions because the relevant modes continue in the adiabatic vacuum. 

\

Then, after the beginning of kination the Bogoliubov coefficients stabilize and the $k$-mode has the simple form
\begin{eqnarray}
\chi_{k}(\eta)= \alpha_k\phi_k(\eta)+
\beta_k \bar{\phi}_k(\eta),\end{eqnarray}
where the adiabatic modes $\phi_k$ has been defined in (\ref{modes})
and,  using the Wronskian $W[f,g]\equiv fg'-f'g$, one has
\begin{eqnarray}
\beta_k=\frac{W[\chi_k,\phi_k]}{W[\bar{\phi}_k,\phi_k]}.
\end{eqnarray}

Another way to perform the calculation
is to use the  justified approximation $\alpha_k(\eta)\cong 1$. Then, the  $\beta$-Bogoliubov coefficient is given by 
\begin{eqnarray}\label{integral}
 \beta_k(\eta)\cong \int_{\eta_i}^{\eta}\omega_k'(\tau)\bar{\phi}_k^2(\tau) d\tau,
\end{eqnarray}
where we have denoted by $\tau_i$ the initial moment when the quantum field is in the vacuum, which occurs at early times during the slow-roll phase because during this era the adiabatic condition $\omega_k'/\omega_k^2\ll 1$ is fulfilled, meaning that there is no particle production. In fact, particles are gravitationally produced during the period between the end  of the slow-roll and the beginning of kination.

\

The numerical calculation of the integral (\ref{integral}) is very oscillating and is very complicated to calculate numerically. For this reason, it seems better to solve the differential equation 
\begin{eqnarray}\label{B}
y_k'-2i\omega_k(\eta)y_k=\frac{\omega'_k(\eta)}{2\omega_k(\eta)},
\end{eqnarray}
with initial condition $y_k(\eta_i)=0$ because,   by taking into account the formula of variation of parameters for first order differential equations,  one has $|\beta_{k}(\eta)|=|y_k(\eta)|$.

\

\begin{figure}[H]
   \centering
   \includegraphics[width=190pt]{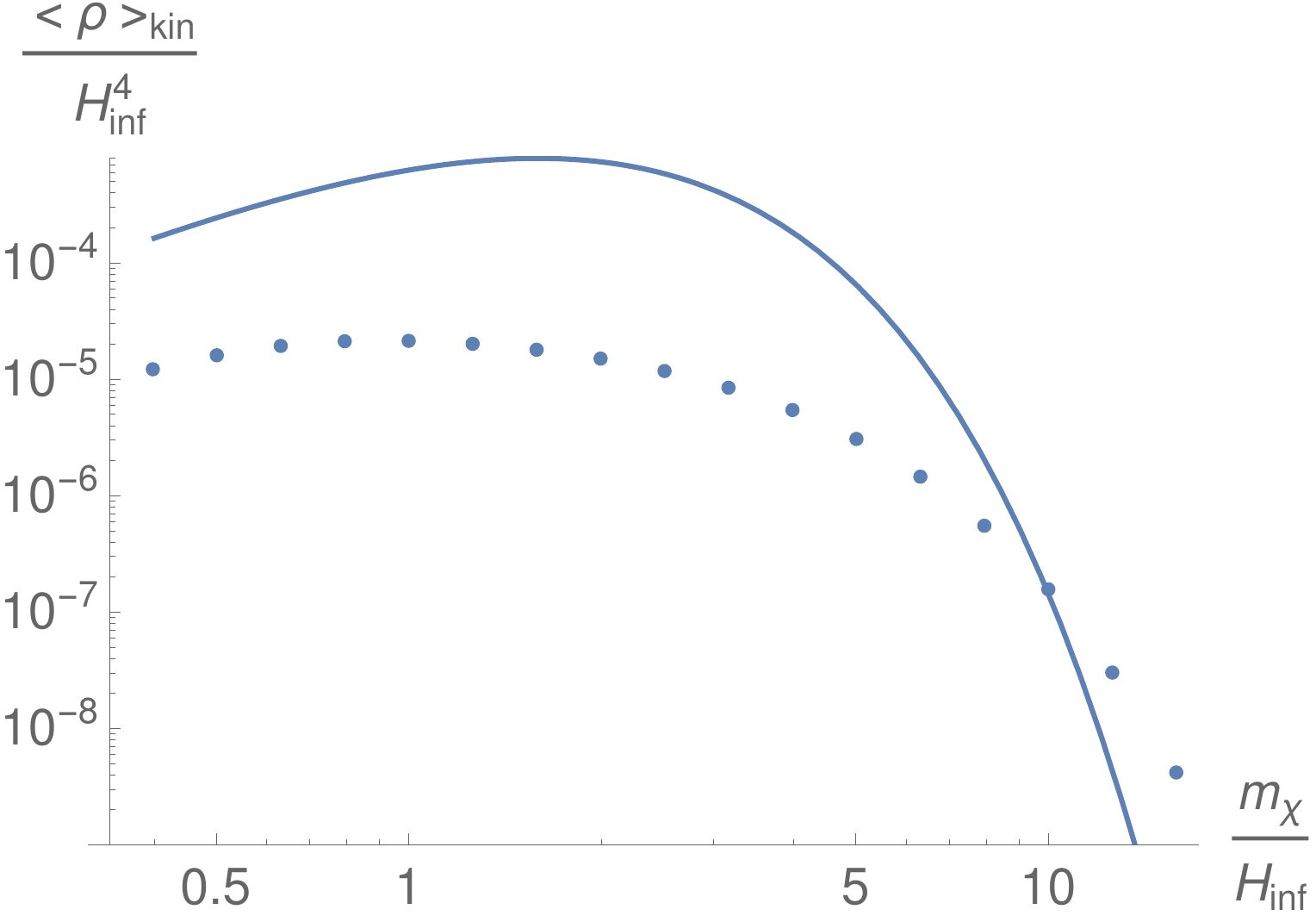}
   \includegraphics[width=190pt]{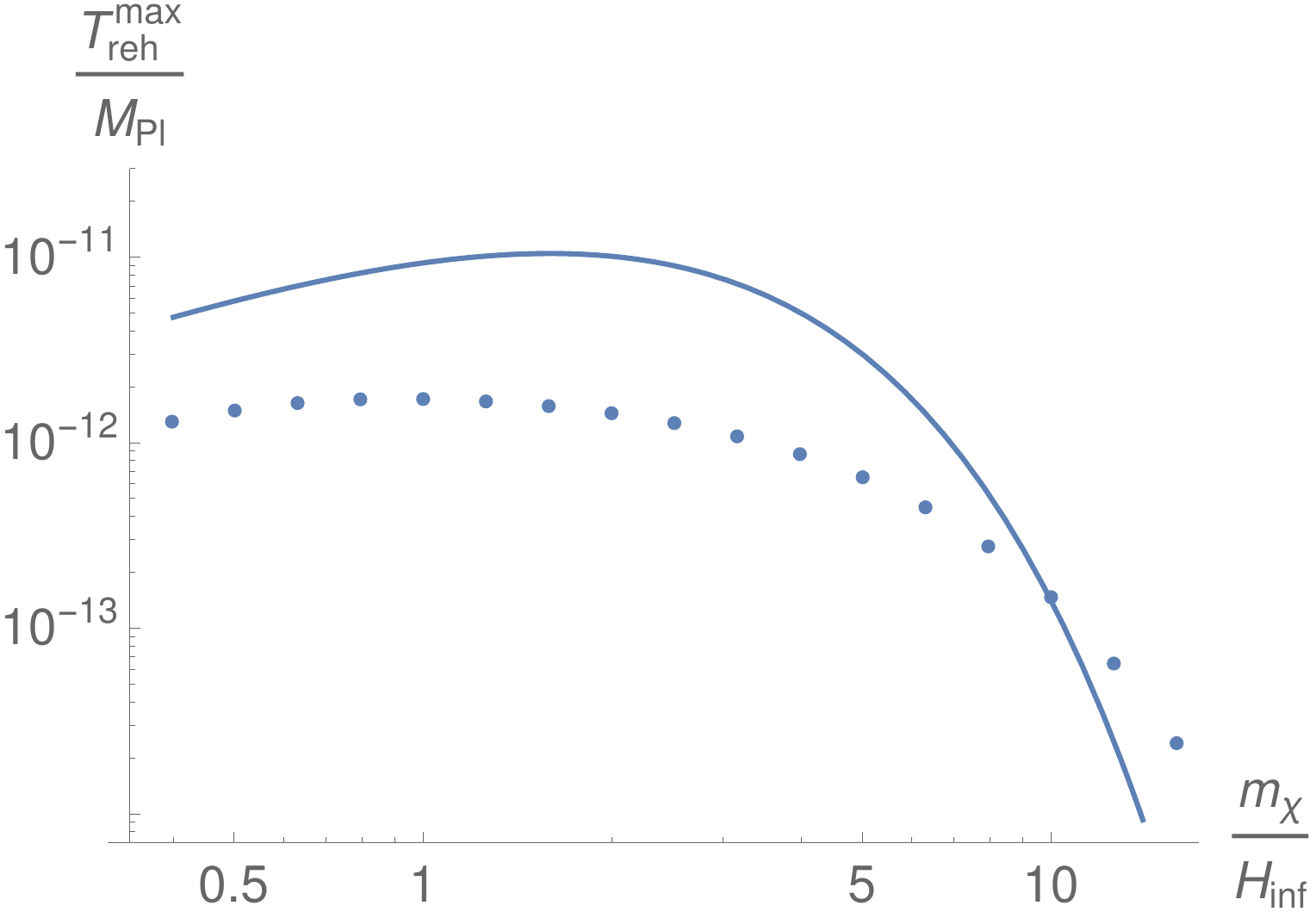}
   \caption{Numerical (in dots) and analytic 
   values for the energy density and the maximum reheating temperature.
   The values of the parameters are $H_{inf}=10^{-6} M_{pl}$, $\alpha=10^{-2}$ and $n=124$.}
   \label{fig:beta}
\end{figure}

{
Finally, it is important to remark the difficulties to perform the numerical calculations taking vacuum initial conditions far from the end of inflation, for example at the horizon crossing (see for a detailed explanation \cite{haro1}), this is why our analytic formula acquires a relevant importance, since it allows to obtain the reheating temperature for all viable models without the need of numerical simulations. In fact, using the Wronskian method and taking vacuum initial condition at a few e-folds before the end of inflation we have obtained that our formula (\ref{rho0}) approximates very well the numerical results. We can see  in Figure  \ref{fig:beta} that   both maximum reheating temperatures, the one calculated analytically and the other numerically, only differ less than an order for the relevant values of the mass $m_{\chi}$.}


\

\section{Conclusions}

In the present work we have found for smooth  non-oscillating backgrounds the   master formula (\ref{master2}), which provides, with good approximation, the energy density of the particles -massive and massless- gravitationally  produced  at the beginning of the kination phase,  for models of Quintessential Inflation. And, consequently, from it one can calculate analytically the reheating temperature for all models  belonging to this class.

\

The formula reproduces the well-known fact that for masses larger than the Hubble rate at the end of inflation the particle production is exponentially suppressed, but for masses of the order of the Hubble rate it leads to very high  reheating temperatures of the order of $10^7$ GeV, which shows that the gravitational particle production mechanism is very efficient for this range of masses. It also reproduces the early well-known results obtained  for light particles non-conformally coupled to gravity. In addition, uising our formula we have shown that the reheating via particle production of heavy particles overcomes the constraints due to  the overproduction of Gravitational Waves during the phase transition from the end of inflation to the beginning of kination.

\

Finally, 
we have also tested our formula with a smooth toy model {and an exponential $\alpha$-attractor} where numerical calculations can be done, showing that our analytic results coincide very well with the numerical ones. {In fact, the theoretical and numerical values of the reheating temperature only differ in one order}. This is a guarantee that our formula can be applied to {other realistic models of Quintessential Inflation} where the numerical calculations have an extreme difficulty. However, checking numerically our analytic results for viable models is a task which deserves future investigations.

\section*{Acknowledgments} 
JdH and LAS are supported by the Spanish grant 
PID2021-123903NB-I00
funded by MCIN/AEI/10.13039/501100011033 and by ``ERDF A way of making Europe''.
JdH
is also supported in part by the Catalan Government 2017-SGR-247. LAS thanks the School of Mathematical Sciences (Queen Mary University of London) for the support provided.

\end{document}